\pdfoutput=1
\documentclass[]{aa}
\usepackage{graphicx}
\usepackage{natbib}

\begin{document}

\title{Mapping of interstellar clouds with infrared light scattered from dust: TMC-1N\thanks{{\it Herschel} is an ESA space observatory with science instruments provided by European-led Principal Investigator consortia and with important participation from NASA.}}

\author{
J. Malinen \inst{1} \and
M. Juvela \inst{1} \and
V.-M. Pelkonen \inst{2,1}\and
M. G. Rawlings \inst{3}
}

\institute{
Department of Physics, University of Helsinki, P.O. Box 64, FI-00014 Helsinki, Finland; johanna.malinen@helsinki.fi
\and
Finnish Centre for Astronomy with ESO, University of Turku, V\"ais\"al\"antie 20, FI-21500 PIIKKI\"O, Finland
\and
National Radio Astronomy Observatory, 520 Edgemont Road, Charlottesville, VA 22903, USA
}


\abstract
{
Mapping of the near-infrared scattered light is a recent method for the study of interstellar clouds, complementing other, more commonly used methods, like dust emission and extinction.
}
{
Our goal is to study the usability of this method on larger scale, and compare the properties of a filamentary structure using infrared scattering and other methods. We also study the radiation field and differences in grain emissivity between diffuse and dense areas.
}
{
We have used scattered near-infrared (NIR) $J$, $H$, and $K$ band surface brightness observations with WFCAM instrument to map a filament TMC-1N in Taurus Molecular Cloud, covering an area of $1^{\circ} \times 1^{\circ}$ corresponding to $\sim$(2.44 pc)$^2$. We have converted the data into an optical depth map and compared the results with NIR extinction and \emph{Herschel} observations of sub-mm dust emission. We have also modelled the filament with 3D radiative transfer calculations of scattered light.
}
{We see the filament in scattered light in all three NIR bands.
We note that our WFCAM observations in TMC-1N show notably lower intensity than previous results in Corona Australis using the same method.
We show that 3D radiative transfer simulations predict similar scattered surface brightness levels as seen in the observations.
However, changing the assumptions about the background can change the results of simulations notably.
We derive emissivity, the ratio of FIR dust emission to column density, by using optical depth in the $J$ band, $\tau_{J}$, obtained from NIR extinction map as an independent tracer of column density. We obtain a value 0.0013 for the ratio $\tau_{250}/\tau_{J}^{Nicer}$.
This leads to opacity or dust emission cross-section
$\sigma_e(250 \mu \rm{m})$ values $1.7-2.4 \times 10^{-25} {\rm cm}^2/{\rm H}$, depending on assumptions of the extinction curve, which can change the results by over 40\%. These values are twice as high as obtained for diffuse areas, at the lower limit of earlier results for denser areas.
}
{
We show that NIR scattering can be a valuable tool in making high resolution maps.
We conclude, however, that NIR scattering observations can be complicated, as the data can show comparatively low-level artefacts. This suggests caution when planning and interpreting the observations.
}

\keywords{ISM: Clouds -- Stars: formation -- Infrared: ISM -- Submillimeter: ISM -- Scattering -- Radiative transfer}

\maketitle

\section{Introduction} \label{sect:intro}

The structure of molecular clouds can be studied via a number of
methods. These include molecular line mapping, observations of dust emission
at far-infrared/sub-millimetre wavelengths, star counts in the optical and
near-infrared (NIR) wavelengths, and measurements of colour excesses of background
stars.

All techniques have their own drawbacks. For example, line and
continuum emission maps are subject to abundance variations (gas and dust,
respectively) and variations in the physical conditions, most notably the
excitation and kinetic temperatures. Mass estimates based on dust emission can also be biased because of line-of-sight temperature variations, especially in high density clouds where star formation is potential~\citep[see, e.g.,][]{Malinen2011}.
The colour excess method provides column
density estimates for extremely narrow lines of sight toward background stars.
However, the intrinsic colours of the stars are usually unknown and this
introduces significant noise, especially at low column densities. A full
extinction map is obtained only after spatial averaging. This means that for
all the listed methods the spatial resolution is usually some tens of arc
seconds or worse. See, e.g.,~\citet{Juvela2006} for a more thorough review of the methods, \citet{Goodman2009} for a comparison of several methods, and \citet{Malinen2012} for a comparison of filament properties derived using NIR extinction and \emph{Herschel} observations of dust emission.

Surface brightness caused by scattered NIR light provides another means of studying cloud structure.
The first observation of scattered NIR light from molecular clouds illuminated by the normal interstellar radiation field (ISRF) was by \citet{Lehtinen1996}. Later the observations of \citet{Nakajima2003} and \citet{Foster2006} (who also named the phenomenon ''cloudshine'') have shown that it is now possible to obtain large maps of the surface brightness of normal interstellar clouds illuminated by the normal ISRF. NIR scattering can therefore be a new, complementary tool for studying the structure of dark clouds. See, e.g.,~\citet{Juvela2006} for a more complete review of the history of scattered light observations in dark clouds.

\citet{Padoan2006} presented a method to determine the cloud column density from the intensity of the near-infrared scattered light. \citet{Juvela2006} analysed the method in more detail using simulations, and developed a method to combine the surface brightness with extinction, to reduce errors caused by wrong assumptions of radiation field or dust properties.
Starting with the known properties of the interstellar dust and ISRF, the papers made predictions for the visibility of
the cloudshine and for the accuracy of the resulting column density estimates.
They also demonstrated, independently of the direct evidence given by the
\citet{Foster2006} data, that such observations are well within the
capabilities of modern wide-field infrared cameras. The main advantage of the new method is the potentially extremely good spatial resolution. When the $J$, $H$, and $K$ bands are used, the method remains accurate in regions with
$A_{\rm V}$ even beyond 10 magnitudes.

The same NIR observations provide data for the
colour excess method, which means that the results of the methods can be compared at
lower spatial resolution. Although both methods depend on near-infrared dust
properties, the main sources of error are different. Comparison of the
results can be used to study the values and variation of near-infrared dust
properties (e.g., albedo and the shape of the extinction curve) and spatial
variations in the strength of the local radiation field. Furthermore,
correlations between wavelength bands give a direct way to estimate the point at which the contribution of dust emission from stochastically heated grains becomes significant. The level of the emission of these so-called Sellgren grains depends on the radiation field~\citep{Sellgren1996}, but is still uncertain in normal interstellar clouds. On the other hand, the amount of the scattered light depends heavily on the grain size~\citep{Steinacker2010}.
These dependencies have implications for models of interstellar dust.

\citet{Juvela2008} used scattered NIR light to derive a column density map of a part of a filament in Corona Australis, and continued the analysis in a larger area in \citet{Juvela2009}. They also compared the NIR data with \emph{Herschel} sub-millimetre data in \citet{Juvela2012b}. \citet{Nakajima2008} applied a similar method to convert NIR scattered light to column density. They used the colour excess of individual background stars to calibrate an empirical relationship between surface brightness and column density, instead of an analytical formula, as used in \citet{Juvela2006,Juvela2008}.
Scattered surface brightness from dense cores in the mid-infrared (MIR) was reported by~\citet{Pagani2010} and~\citet{Steinacker2010}. \citet{Steinacker2010} named this phenomenon ``coreshine'' as a counterpart to ``cloudshine'', which is also observed in the outer parts of the clouds.

In this paper, we study a filament in the Taurus molecular cloud using observations of NIR light of a $1^{\circ} \times 1^{\circ}$ field observed with WFCAM instrument~\citep{Casali2007}. Distance to the Taurus molecular cloud is $\sim$140 pc, making it one of the closest relatively high latitude clouds, and consequently one of the most studied star-forming regions~\citep[see, e.g.,][]{Cambresy1999,Nutter2008,Kirk2012,Palmeirim2013}.
In \citet{Malinen2012}, we compared the properties of this filament derived using NIR extinction and dust emission observed with \emph{Herschel}. 
Here, we construct maps of the diffuse surface brightness, determine the intensity of the NIR scattered light, and derive the optical depth based on scattered NIR light using the method presented in \citet{Padoan2006}.
We compare the scattered light images with the other tracers, NIR extinction and sub-millimetre dust emission and, using these results, draw some conclusions regarding the intensity and spectrum of the local ISRF. We also perform radiative transfer modelling to compare observations with the level of NIR and MIR scattered light that is expected using standard ISRF levels and standard dust models.

The contents of the article are the following:
We present observations and data processing in Sect.~\ref{sect:observations}. We describe the method for deriving optical depth from scattered NIR surface brightness in Sect.~\ref{sect:method}. We derive NIR surface brightness maps and optical depth maps based on observations of dust emission, NIR extinction, and NIR scattered surface brightness and compare the results in Sect.~\ref{sect:results}. We describe radiative transfer modelling of a filament seen in scattered light in Sect.~\ref{sect:models}. We discuss the results in Sect.~\ref{sect:discussion} and present our conclusions in Sect.~\ref{sect:conclusions}.

\section{Observations and data processing} \label{sect:observations}

\subsection{WFCAM}

We have used the Wide Field CAMera (WFCAM) \citep{Casali2007} of the United Kingdom InfraRed Telescope (UKIRT) to observe a $1^{\circ} \times 1^{\circ}$ field in the NIR $J$, $H$, and $K$ bands (1.25, 1.65, and 2.22 $\mu m$, respectively). The field, which we call TMC-1N~\citep{Malinen2012}, is in the Taurus molecular cloud complex north of TMC-1. The central coordinates of this field are RA~(J2000) 4h39m36s and Dec~(J2000) +26$^{\circ}$39$'$32$''$. At a distance of 140 pc, this corresponds to an area of $\sim$(2.44 pc)$^2$. 
The Galactic latitude of this area is approximately $-13.3^{\circ}$.
The target field was chosen based on the Taurus extinction maps of \citet{Cambresy1999} and \citet{Padoan2002}. 
According to~\citet{Rebull2011}, there are no strong young stellar objects (YSOs) in TMC-1N that could cause additional scattered light and therefore complicate the analysis. There is mainly just one continuous filament in the field.
The $A_{\rm V}$ range of the area is suitable for our method: there are some regions with $A_{\rm V}\sim20^{\rm m}$, but the mean value remains well below 10 magnitudes, even in most parts of the filament.

We applied sky correction using offset fields to be able to measure faint surface brightness features.
The observations were made during 13 nights between 2006--2008 
using 2 $\times$ 2 pointings towards the selected field. Because WFCAM consists of four separate CCD arrays, the result is a $1^{\circ} \times 1^{\circ}$  image
consisting of 4 $\times$ 4 subimages. We used four separate OFF fields.
The standard data reduction was conducted in accordance with the normal pipeline routine\footnote{\tt http://casu.ast.cam.ac.uk/surveys-projects/wfcam/technical}, including, e.g., dark-correction, flatfielding (including internal gain correction), decurtaining, sky correction, and cross-talk. 
Standard decurtaining methods\footnote{\tt http://casu.ast.cam.ac.uk/surveys-projects/wfcam/\\technical/decurtaining} were needed to remove the stripes caused by the instrument. The details of the observations are shown in \citet{Malinen2012} where we compared the filament properties derived using NIR extinction and sub-millimetre dust emission observed with \emph{Herschel}. There, the NIR data were calibrated to magnitudes with the help of 2MASS catalogue stars to derive an extinction map with the NICER method \citep{Lombardi2001}. Here, in order to study the surface brightness, we calibrated the data from magnitudes to MJy/sr units using aperture photometry of several stars in each frame.

The reduced images contained residual gradients in each sub-image, shown as brightening of the signal towards the frame edges. 
These are probably of instrumental origin. The size of the gradients as a percentage of overall sky level varies, but is typically between 0.002--0.008. \citet{Dye2006} and \citet{Warren2007} report several types of artefacts in WFCAM data and note that removing sky subtraction residuals is a complex problem, especially when observing near the Moon. However, during our observations the Moon was 
always further than 40 degrees away. In addition, the moonlight artefacts that cause most problems are local scattered light from dust on the optics, not gradients. The master twilight flats may have low level gradients present in them, given the size of the WFCAM field-of-view. The dark correction can also leave low level reset anomalies, particularly near the detector edges. It is possible that our data are showing comparatively low-level residual issues that the general surveys have not encountered.

We first removed the stars using the Iraf Daophot-package, masked the remaining bad pixels (noisy borders and residuals of bright or saturated stars) and performed the surface brightness calibration. We modelled the gradients with a method described in the following subsection (Sect.~\ref{sect:gradrem}). 
We convolved the calibrated frames with a 2$''$ Gaussian beam and resampled the data onto 0.8$''$ pixels. The obtained gradients were then subtracted from the frames. The maximum intensity of the filament is a few times higher than the typical magnitude of the instrumental gradients.

We further median-filtered the images to $\sim$16.8$''$ resolution to diminish the effect of residual stars and stellar artefacts, and finally convolved them to $\sim$40$''$ resolution, for later comparison with \emph{Herschel} observations at that resolution.
We combined the frames to a full map using Montage\footnote{\tt http://montage.ipac.caltech.edu/} and resampled the data onto 8$''$ pixels for the analysis. For comparisons with NICER data, we further convolved the intensity maps to 60$''$ resolution.
We subtracted the background from the maps, using an area of low column density outside the filament as a reference, see Fig.~\ref{fig:tau_fullmaps} (middle frame).

A map of visual extinction, $A_V$, of the TMC-1N area was already derived and presented in~\citet{Malinen2012}, using a~\citet{Cardelli1989} extinction curve with $R_V = 4.0$. In this paper, we mainly use the $J$ band optical depth $\tau_J$, which is in practice independent of the assumption of the $R_V$ value.
For comparison, a \citet{Cardelli1989} extinction curve with $R_V = 4.0$ gives the relation $\tau_J = 0.2844 \times A_V$.

\citet{Roy2013} discuss the effect of the finite width of the filters on the observed extinction. In their Fig. 10, they show the relation of colour excess $E(J-K_S)$ obtained with 2MASS filters to the corresponding monochromatic colour excess as a function of $N_H$. In TMC-1N, having an $A_V$ of mostly less than 20$^m$, the effect is small (less than 5\% even in the densest parts), and we have not made this correction.

\subsubsection{Gradient modelling}  \label{sect:gradrem}

During the data reduction, separate analysis was carried out to model and subtract the residual artefacts of the surface brightness data.
To be able to model the large scale gradients, we first filtered the masked and calibrated data with a median filter to $\sim$40$''$ resolution. The gradients were strongest near the borders, which meant that the overlap between the frames could not be reliably used to aid the gradient removal.
To model the gradient, therefore, we assumed a similar relation between the column density and the surface brightness as the one described in Sect.~\ref{sect:method}. We used the \emph{Herschel} column density map described in Sect.~\ref{sect:herschel} as an independent tracer of the filament structure in the fitting of the gradients. In principle, the extinction map obtained from NIR reddening of background stars could also have been used, but we chose \emph{Herschel} data because of its lower noise.

We modelled the gradients with a third order surface describing the artificial gradient in the image, with an additional column density dependent term. We used least squares fitting to minimize, separately for each band $i$ in ($J$, $H$, $K$) and each frame, the residual
\begin{equation}
Res = P + a_i(1-e^{-b_iN}) - I_i
\label{eq:res}
\end{equation}
where $P$ is the complete third order polynomial for a 2-dimensional surface (10 terms), $I_i$ is the observed intensity map in band $i$, and $N$ the column density derived from \emph{Herschel} data.

The parameter $b$ describes the saturation of the relation between $I$ and $N$, and it mainly depends on the grain properties~\citep{Juvela2006}. If we presume that the dust grains in the cloud are similar as in the model of~\citet{Juvela2006}, the same $b$ values should apply as long as the optical depth is not high.
Therefore, we used the parameter values $b_J$ = 0.34, $b_H$ = 0.23, and $b_K$ = 0.15, (in 1/mag units, as the model used $A_V$ values instead of $N$) taken from the model data. The parameter $b$ does not depend on the strength of the incoming radiation, but in some amount on the direction distribution. The model was based on a large cloud with an inhomogeneous density distribution and an isotropic radiation field. TMC-1N, on the other hand, has a single, filamentary structure, and a larger part of the radiation comes in from the observer's side. We do not expect the model to describe the filament perfectly, but the model parameter values served as a good starting point.

The coefficients of the polynomial are free parameters and are used to describe the artificial gradients that will be removed in the subsequent analysis. The parameters $a_i$ are related to the actual signal from the cloud and are the same for all frames in the same band. They are also kept as free parameters to minimise the effect the cloud structure has on the gradient fits. The exponential term $M = a_i(1-e^{-b_iN})$ is only an approximation of the column density dependence of the scattered light. However, because gradients are described only as third order surfaces over each frame and scattered light is visible only in some parts of each frame, the obtained gradient model is not expected to be strongly dependent on the reference map, here the column density derived from \emph{Herschel}. In particular, the scaling between column density and the surface brightness is a free parameter and thus no assumption is made of the expected level of NIR surface brightness. Furthermore, the non-linear approximation represented by the term $M$ becomes important only at $A_V\sim10^m$.
To obtain a better fit for the central part of the frames used in the following analysis, the gradients were fitted excluding a 10\% wide border in each frame. For comparison, we also performed a fit to the whole area of the frames, in order to better fit the borders of the frames.
The fitting provides the model $P$ (for each frame and band) of the gradients that we subtract from the original surface brightness images. Note that the exponential term $M$ is not subtracted as that is not part of the artefact.

We tested the reliability of the gradient removal method in several ways.
We calculated the residual between the used model and the corrected surface brightness map, $S = I - P$, with the function
$Res = a_i(1-e^{-b_iN}) - S_i$, where $b_i$ are the constants used in the gradient modelling and $a_i$ are the values obtained from the fit. $N$ is the map used in the gradient modelling. We used surface brightness maps in $\sim$2.2$\arcsec$ resolution.
The mean values for the residual are -0.0052, 0.00035, and 0.000074 MJy/sr for the $J$, $H$, and $K$ bands, respectively. The standard deviation for the residuals of the same bands are 0.017, 0.031, and 0.021 MJy/sr. Noise, especially near the borders, increases the obtained standard deviation values. The absolute values for the mean residuals for all bands are less than 0.006 MJy/sr, indicating that the minimisation of the residual in Eq.~\ref{eq:res} has been effective. The filament is not apparent in the residuals nor in the removed gradients.

We also compared the surface brightness maps derived using different areas for the fitting of the gradient, full area or excluding 10\% wide borders of each frame. After the gradient removal, we compared the surface brightness maps (in 40$\arcsec$ resolution) by calculating the difference between the maps in the final masked area shown in Fig.~\ref{fig:I_J_masked}. The means of the difference of the final masked area are 0.0026, 0.0026, and -0.0014 MJy/sr for the $J$, $H$, and $K$ bands, respectively. The standard deviations for the difference of the same bands are 0.0044, 0.0072, 0.0052 MJy/sr. This indicates that small differences in the fitting area do not cause a significant difference in the central parts of the corrected frames. Near the borders of the frames, the difference can be larger.

\subsection{Herschel}  \label{sect:herschel}

The Taurus molecular cloud has been mapped with \emph{Herschel}~\citep{Pilbratt2010} as part of the Gould Belt Survey\footnote{\tt http://gouldbelt-herschel.cea.fr}~\citep{Andre2010}, see~\citet{Kirk2012} and ~\citet{Palmeirim2013}.
We used SPIRE~\citep{Griffin2010} 250\,$\mu$m, 350\,$\mu$m, and 500\,$\mu$m maps of the TMC-1N field. 
We obtained the data from the \emph{Herschel} Gould Belt Survey consortium.
The observation identifiers of the data are 1342202252 and 1342202253.
The details of the \emph{Herschel} data and the derivation of column density map are presented in~\citet{Malinen2012}.

For easier comparison with other maps and to avoid the assumptions needed to make column density maps, we used optical depth $\tau_{250}$ instead. We use the notation $\tau_{250}$ to mean optical depth $\tau_{\nu}$ at wavelength 250\,$\mu$m ($\nu = $1200GHz).
We derived  $\tau_{250}$ maps from \emph{Herschel} intensity and colour temperature maps, using a value of 1.8 for the spectral index $\beta$. Recent studies with \emph{Planck}~\citep{Planck2011b} show that this is a good estimate for Taurus.
For comparison, we also made the analysis with the commonly used value of $\beta = 2.0$.

\subsection{Spitzer}

The Taurus filament is covered by observations of the Spitzer IRAC
instrument~\citep{Fazio2004}.
The IRAC data (observation numbers 11230976 and 11234816) are
from the Taurus Spitzer legacy project (PI D. Padgett). The archival
IRAC maps show some checkered pattern. We therefore started with the
pipeline-reduced, artefact-mitigated images (cbsd images). The
residual offsets were corrected by subtracting the median value from each frame and by estimating the residual offsets with respect to the
median of all frames \footnote{see Spitzer Data Analysis Cookbook,
Sect. 4.4, at
\tt http://irsa.ipac.caltech.edu/data/SPITZER/docs/\\
dataanalysistools/cookbook/}.
The final mosaic image was then made with the MOPEX tool~\citep{Makovoz2006}.

\section{Deriving optical depth from scattered NIR surface brightness} \label{sect:method}

\citet{Padoan2006} presented a method for converting scattered surface brightness into column density. \citet{Juvela2006,Juvela2008} discussed the method and its reliability in more detail. We review the main points of the method given in these articles and formulate it to suit this study.

Based on the one-dimensional radiative transfer equation in a homogeneous medium, the relationship between surface brightness ($I$) and column density ($N$) can be approximated with the formula
\begin{equation}
I = a(1-e^{-bN})
\label{eq:I_N}
\end{equation}
where the parameters $a$ and $b$ are positive constants defined separately for each band. The parameter $b$ scales the column density to optical depth, $\tau$.
The function depicts the radiation coming from the cloud. We presume that the observed radiation is mainly caused by scattering.
Also other tracers of cloud structure, such as extinction $A_V$ or optical depth $\tau$ can be used instead of column density $N$. In that case, the value and unit of the parameter $b$ must be changed accordingly.
The parameters $b$ mainly depend on the NIR dust properties and the parameters $a$ on the radiation field, although the interpretation is not this straightforward~\citep{Juvela2006}. Equation~\ref{eq:I_N} is exactly valid only for a homogeneous cloud. In real clouds it works only as an approximation. It is expected that both parameters change if the dust properties, radiation field or cloud geometry change. Both can, however, be treated just as empirical parameters.

In areas of low optical depth, the intensity of NIR scattered light is directly proportional to the column density as shown in the relation
\begin{equation}
I = abN.
\label{eq:I_N_lin}
\end{equation}
The product $ab$ gives the scattered intensity per column density. With higher column densities ($A_V \sim 10^m$), the NIR intensity values begin to saturate, starting from the shorter wavelengths. Thus, the relation becomes more non-linear with increasing column density.

Equation~\ref{eq:I_N} can be considered as an empirical description of the relation between surface brightness and column density, and it is not necessarily the optimal functional form to describe this relation. However, as long as the data points follow this relation, it is possible to derive approximations for column density. Based on numerical simulations~\citep{Padoan2006,Juvela2006}, this function is rather reliable in areas of low to medium visual extinction ($A_V = 1-15^m$), where the saturation is not strong. 
If no independent column density estimates are available, only the ratios between different bands can be derived from surface brightness observations. However, NIR observations also provide the colour excesses of background stars, and an extinction map can therefore be calculated and used to derive the necessary parameter values~\citep{Juvela2006}.

To derive an optical depth $\tau_{J}^{SB}$ map from the surface brightness maps, we obtain $a$ and $b$ parameter values for each band from $I_{\nu}/\tau_{J}^{Nicer}$ correlations by fitting Eq.~\ref{eq:I_N} to the data, using optical depth $\tau_J^{Nicer}$ instead of column density $N$ in the exponential part.
Thus, Eq.~\ref{eq:I_N} is used to establish a non-linear scaling between the extinction that increases linearly with column density and the surface brightness for which the increase is non-linear. Once the parameters of this mapping have been determined, we can convert surface brightness observations directly to estimates of column density, possibly at a resolution higher than what is possible using background stars alone.
Using the obtained parameter values, we calculate a value for each pixel of the $\tau_{J}^{SB}$ map by minimising the squared sum of residuals 
\begin{equation}
R_{\rm pixel}^2 = \sum\limits_{i=J,H,K}(I_i-M_i)_{\rm pixel}^2
\label{eq:tauSB_1}
\end{equation}
where $I_i$ are the pixel values of the intensity maps, and 
\begin{equation}
M_i = a_i(1.0-e^{-b_i\tau_J^{SB}})
\label{eq:tauSB_2}
\end{equation}
for each $i$ in ($J$, $H$, $K$). Minimisation is needed, as the observed intensities do not follow Eq.~\ref{eq:I_N} exactly due to noise and possible model errors.

\section{Results} \label{sect:results}

\begin{figure*}
\includegraphics[width=6cm]{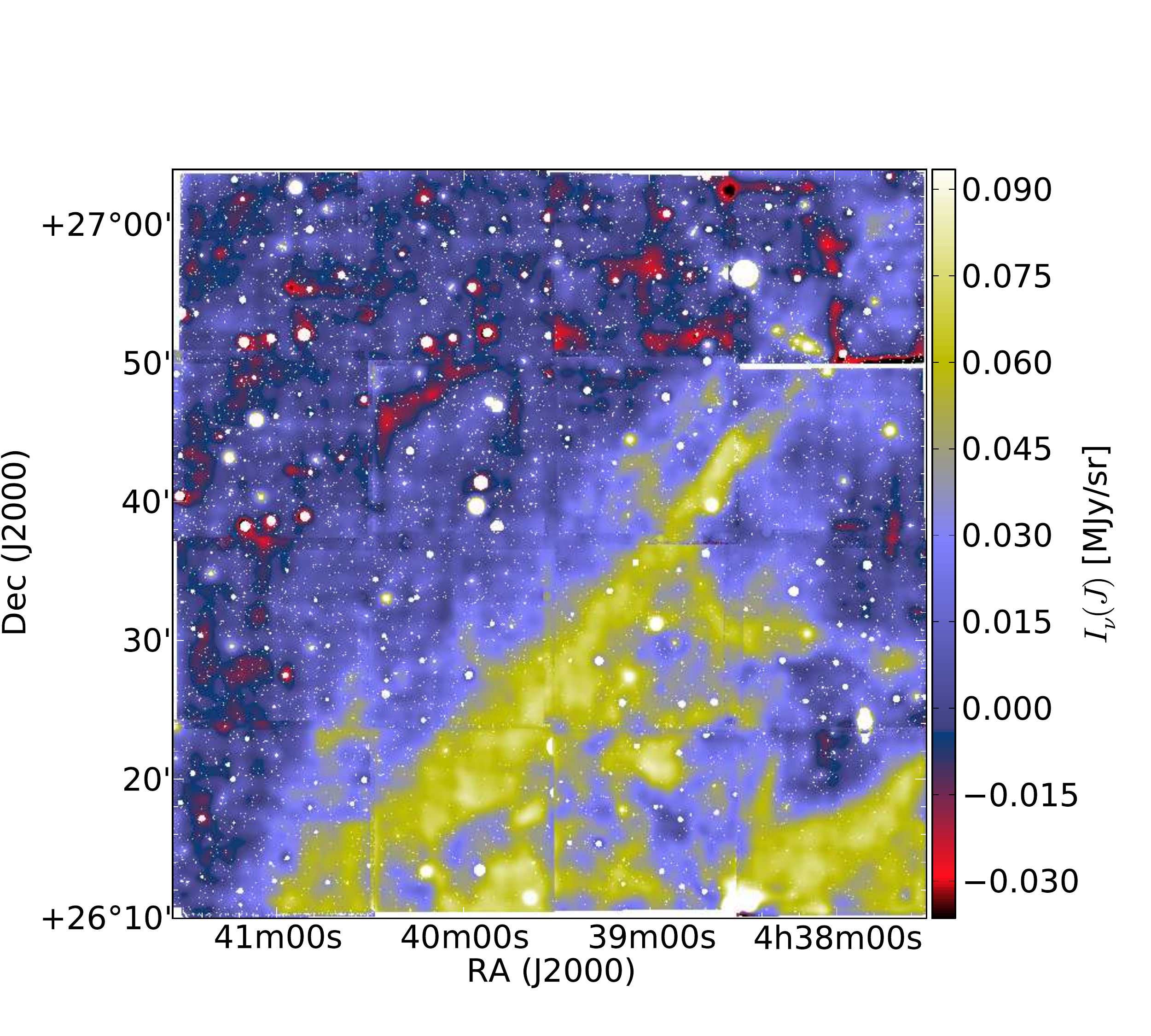}
\includegraphics[width=6cm]{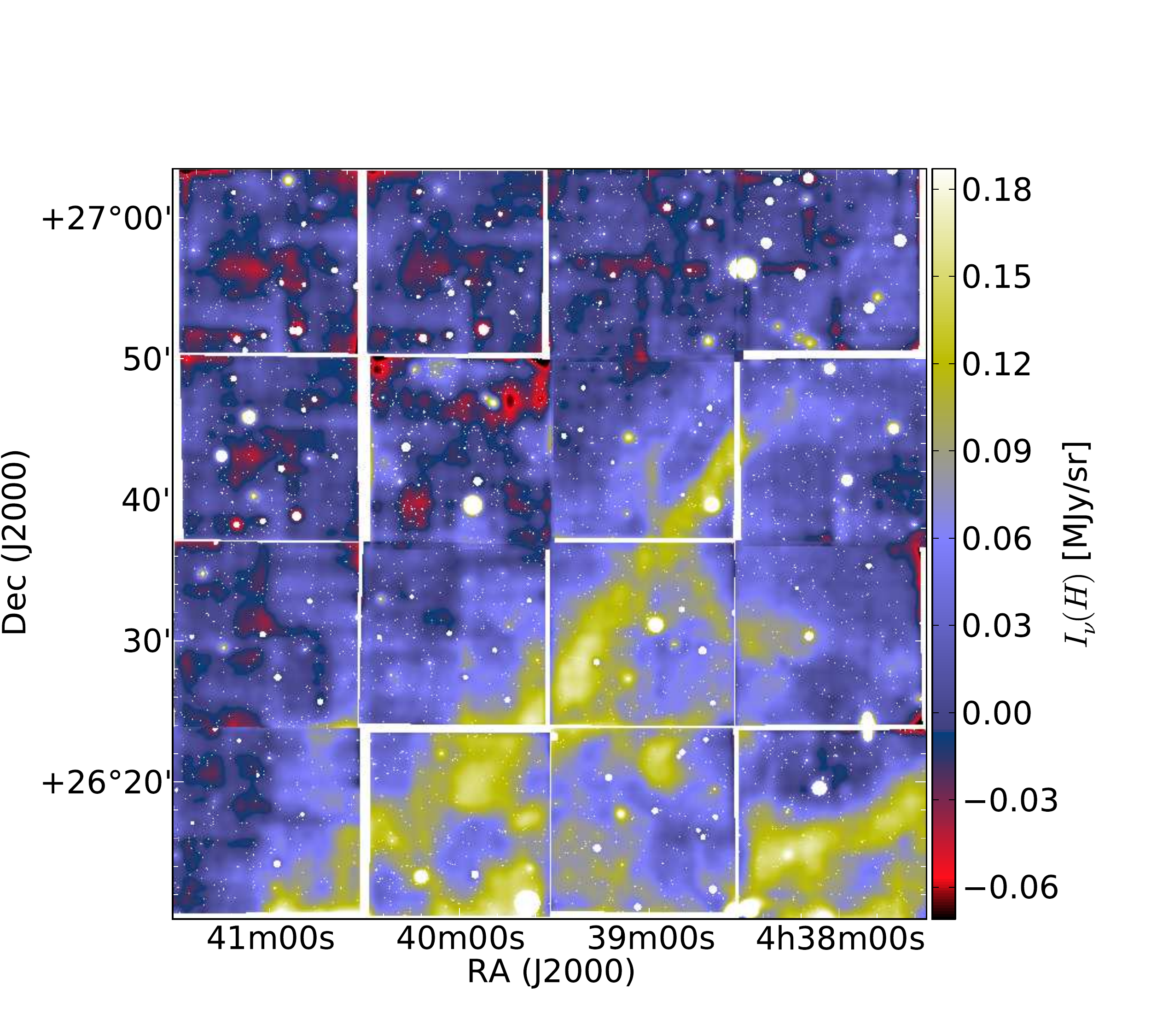}
\includegraphics[width=6cm]{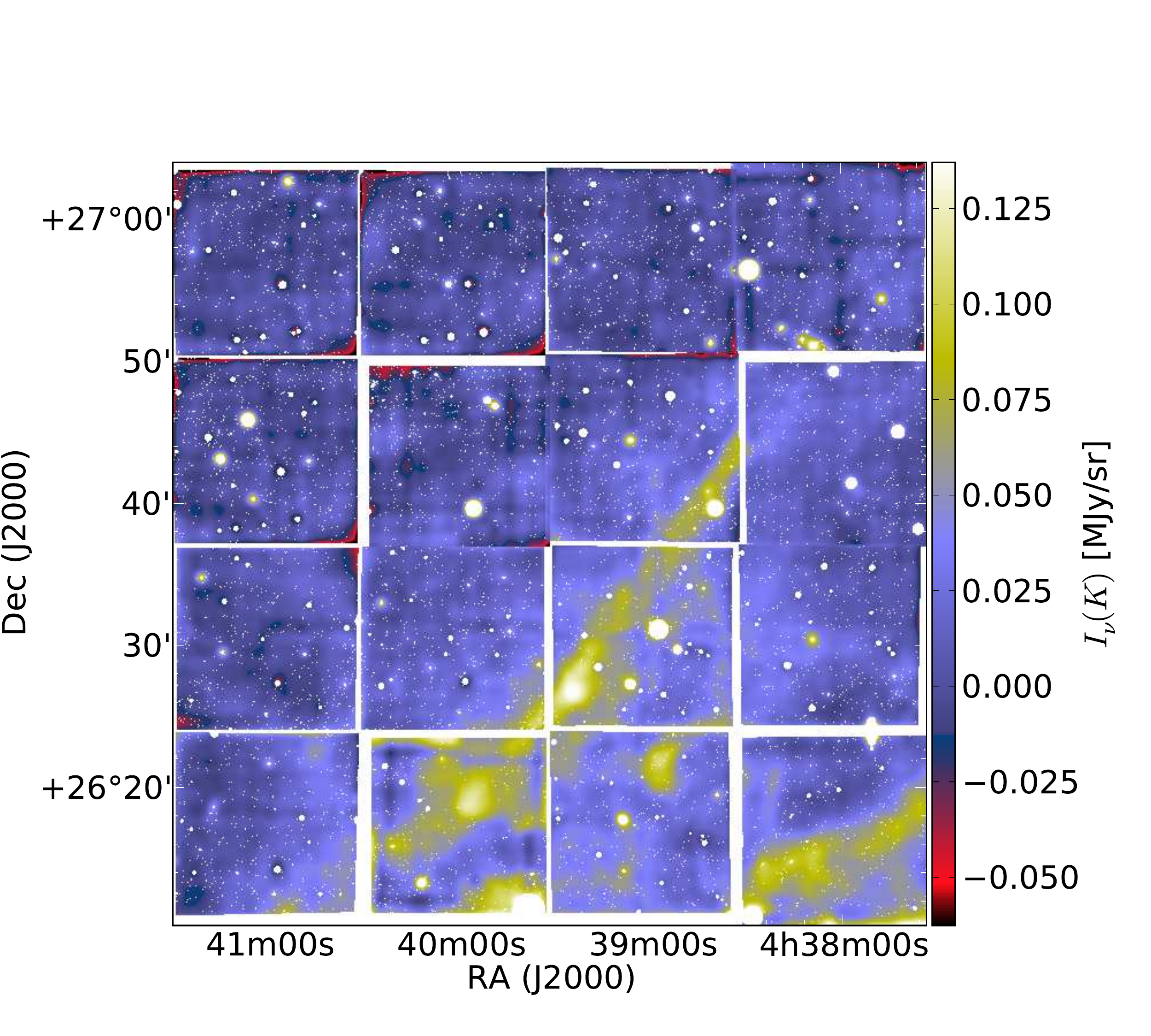}
\caption{$J$, $H$, and $K$ band intensity maps observed with WFCAM and convolved to 40$''$ resolution. Masked pixels are marked white.}
\label{fig:I_J}
\end{figure*}

\begin{figure}
\includegraphics[width=9cm]{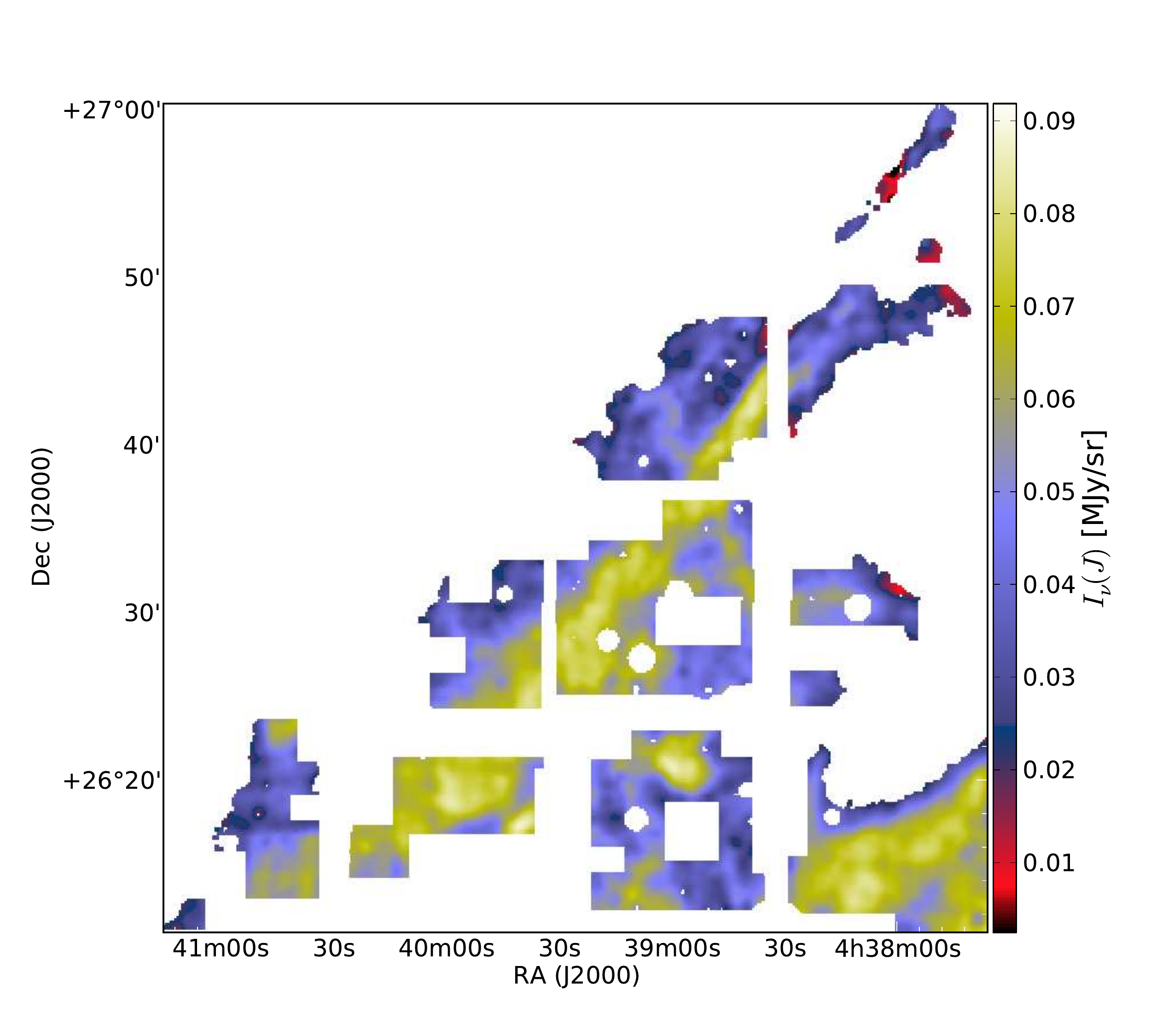}
\caption{J band intensity map with low column density areas (below 1.0$\times10^{21}$cm$^{-2}$) and bad pixels masked. Only the shown unmasked data are used in the scatter plot analysis.}
\label{fig:I_J_masked}
\end{figure}

Surface brightness map in $J$ band with 40$''$ resolution is shown in Fig.~\ref{fig:I_J}. In this figure, the gradient removal is made using a fit to the whole area of each frame.
In the analysis, bad pixels, border regions suffering from imperfect gradient removal, and the main artefacts remaining after star removal were all masked.
We also limited the analysis to areas where \emph{Herschel} column density values are above 1.0$\times10^{21}$cm$^{-2}$. The masked $J$ map used in the analysis is shown in Fig.~\ref{fig:I_J_masked}. We used the same mask for all maps ($J$, $H$, $K$, NICER, and \emph{Herschel}) in the analysis. 
The reference area used for background subtraction is shown in Fig.~\ref{fig:tau_fullmaps} (middle frame). The median $\tau_J$ in this area is $\sim$0.1.

\begin{figure}
\includegraphics[width=8cm]{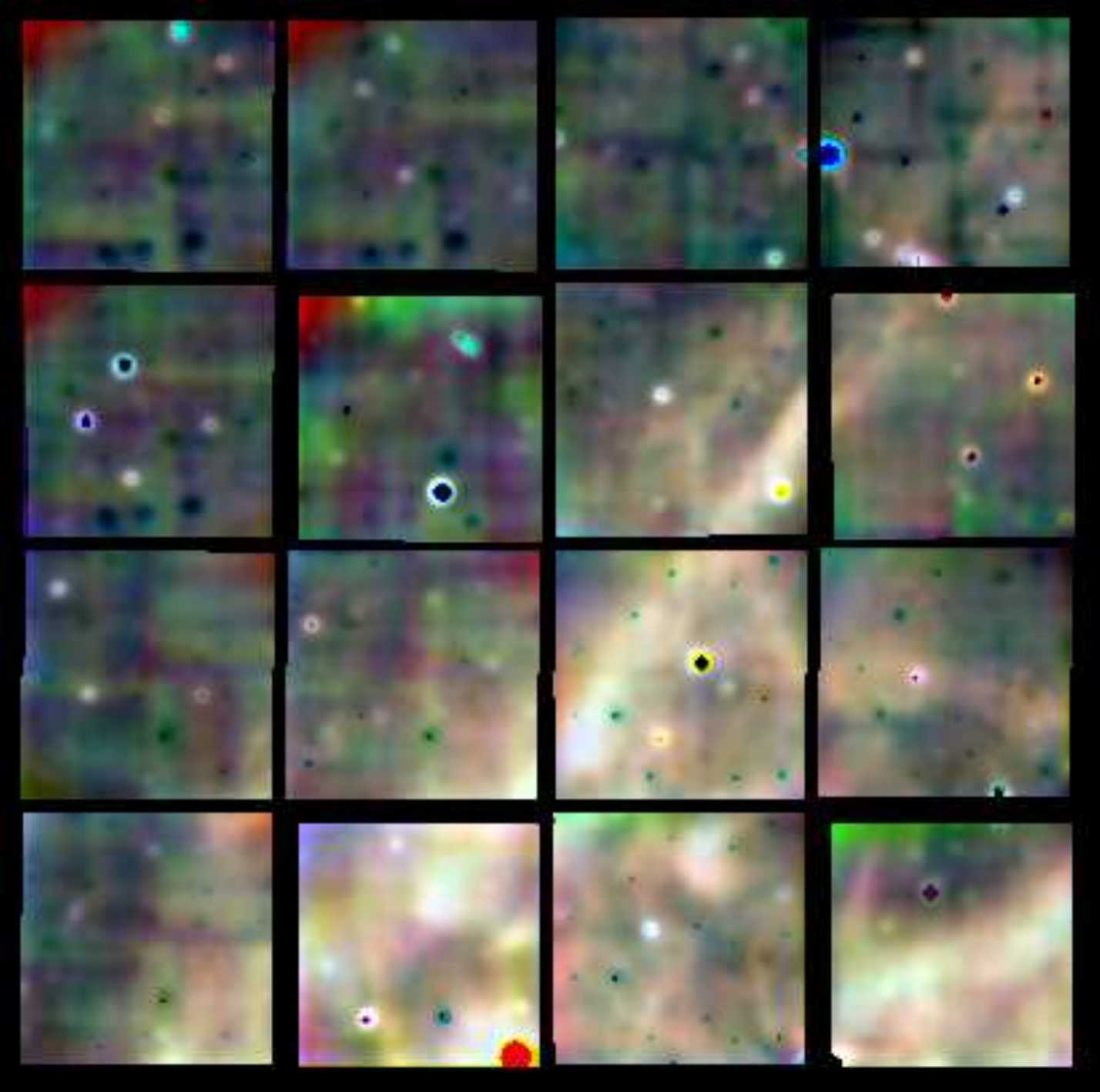}
\caption{Combined three-colour image of WFCAM $J$, $H$, and $K$ band intensity maps with 40$''$ resolution. $J$, $H$, and $K$ bands are encoded in red, green, and blue, respectively. The removed gradients were fitted to the frames excluding 10\% wide borders of each frame.}
\label{fig:RBG}
\end{figure}

A combined three-colour image of WFCAM $J$, $H$, and $K$ band intensity maps with 40$''$ resolution is shown in Fig.~\ref{fig:RBG}. 
Here, the removed gradients were fitted to the frames excluding 10\% wide borders of each frame.
The remaining gradients can most clearly be seen near the corners of the frames. Borders of the frames have been masked.

\subsection{Correlations between NIR surface brightness and optical depth derived from extinction and dust emission} \label{sect:scatterplot}

\begin{figure}
\centering
\includegraphics[width=9cm]{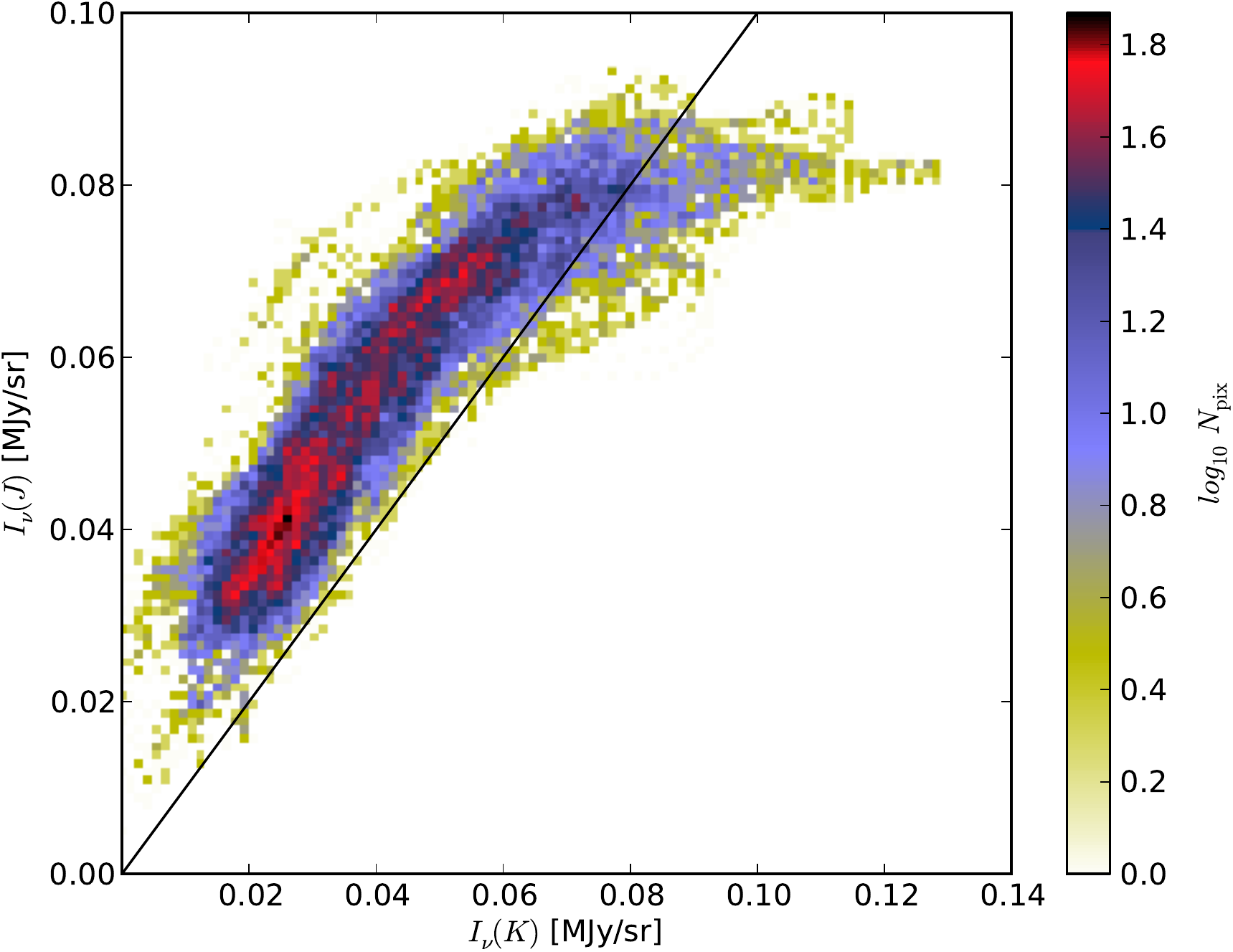}
\includegraphics[width=9cm]{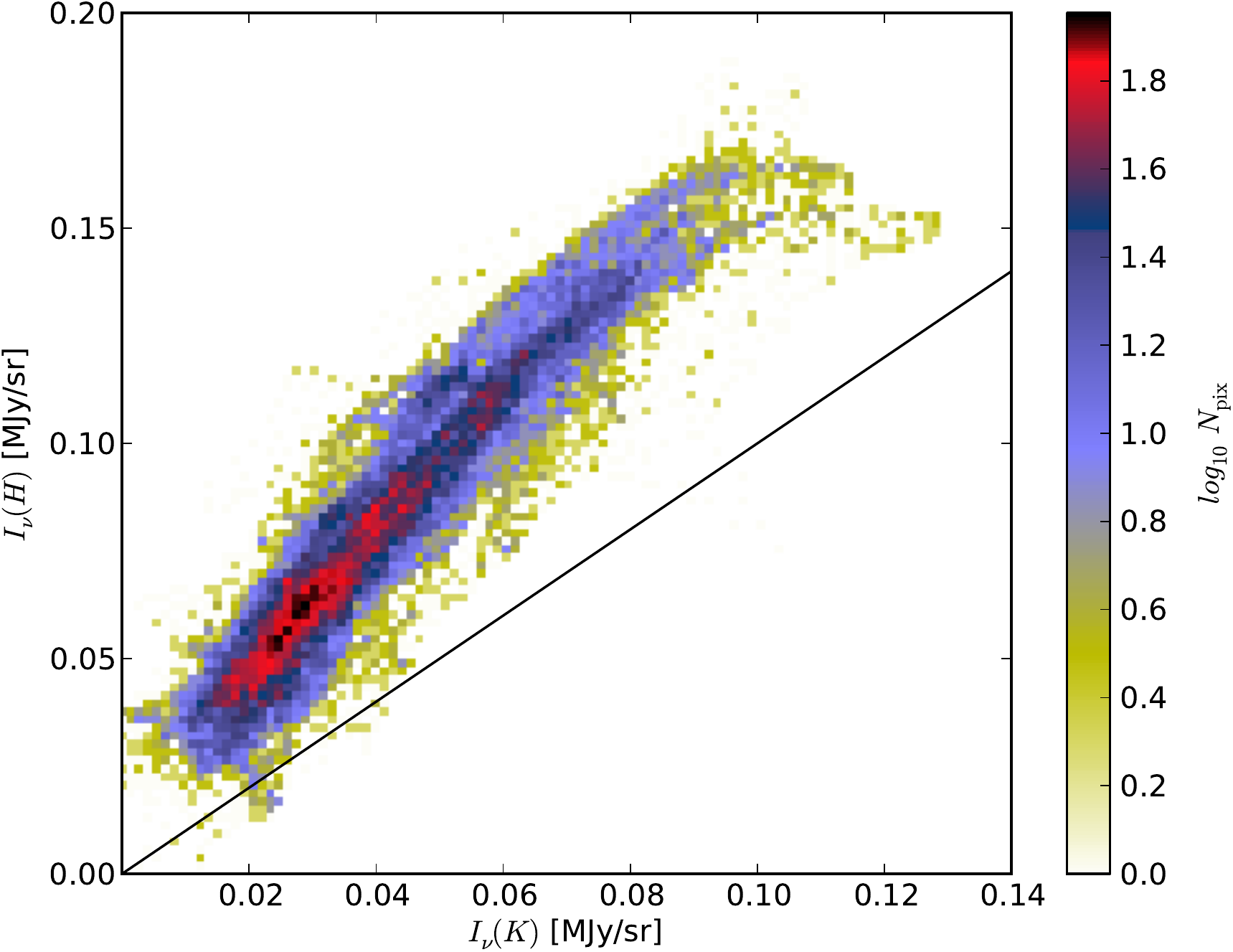}
\caption{
Surface brightness in $J$ band, $I_J$, as a function of the $K$ band, $I_K$, (upper frame) and $I_H$ as a function of $I_K$ (lower frame) for the main filament area shown in Fig.~\ref{fig:I_J_masked}. The $y = x$ function is denoted by a black line. The colour scale of the 2D histogram indicates the logarithmic density of the points, both here and in other figures. The data are convolved to 40$''$ resolution.}
\label{fig:scpl_filament_I_I}
\end{figure}

We show correlations between the NIR surface brightness in $J$, $H$, and $K$ bands for the main filament area in Fig.~\ref{fig:scpl_filament_I_I}. 
The main cloud of points is concentrated to values below 0.08 MJy/sr in $I_J$, 0.17 MJy/sr in $I_H$, and 0.12 MJy/sr in $I_K$. The correlation between $I_J$ and $I_K$ is approximately linear up to $I_K\sim$0.06 MJy/sr, after which $I_J$ starts to saturate. Similarly, the correlation between $I_H$ and $I_K$ is approximately linear up to $I_K\sim$0.08 MJy/sr, after which $I_H$ starts to saturate.

Compared to Fig. 3 in Juvela et al. (2008) for Corona Australis, our WFCAM data for TMC-1N show similar linear relations between $I_J$ and $I_K$ and between $I_H$ and $I_K$ in the low end of the intensity scale, although our data show slightly lower values.
In TMC-1N, the main cloud of points is concentrated to $I_K$ values below 0.12 MJy/sr, whereas in Corona Australis there are also points up to $I_K\sim$0.7 MJy/sr. In Corona Australis, the correlations turn to negative between $I_K\sim$0.4-0.7 MJy/sr.

\begin{figure*}
\centering
\includegraphics[width=6cm]{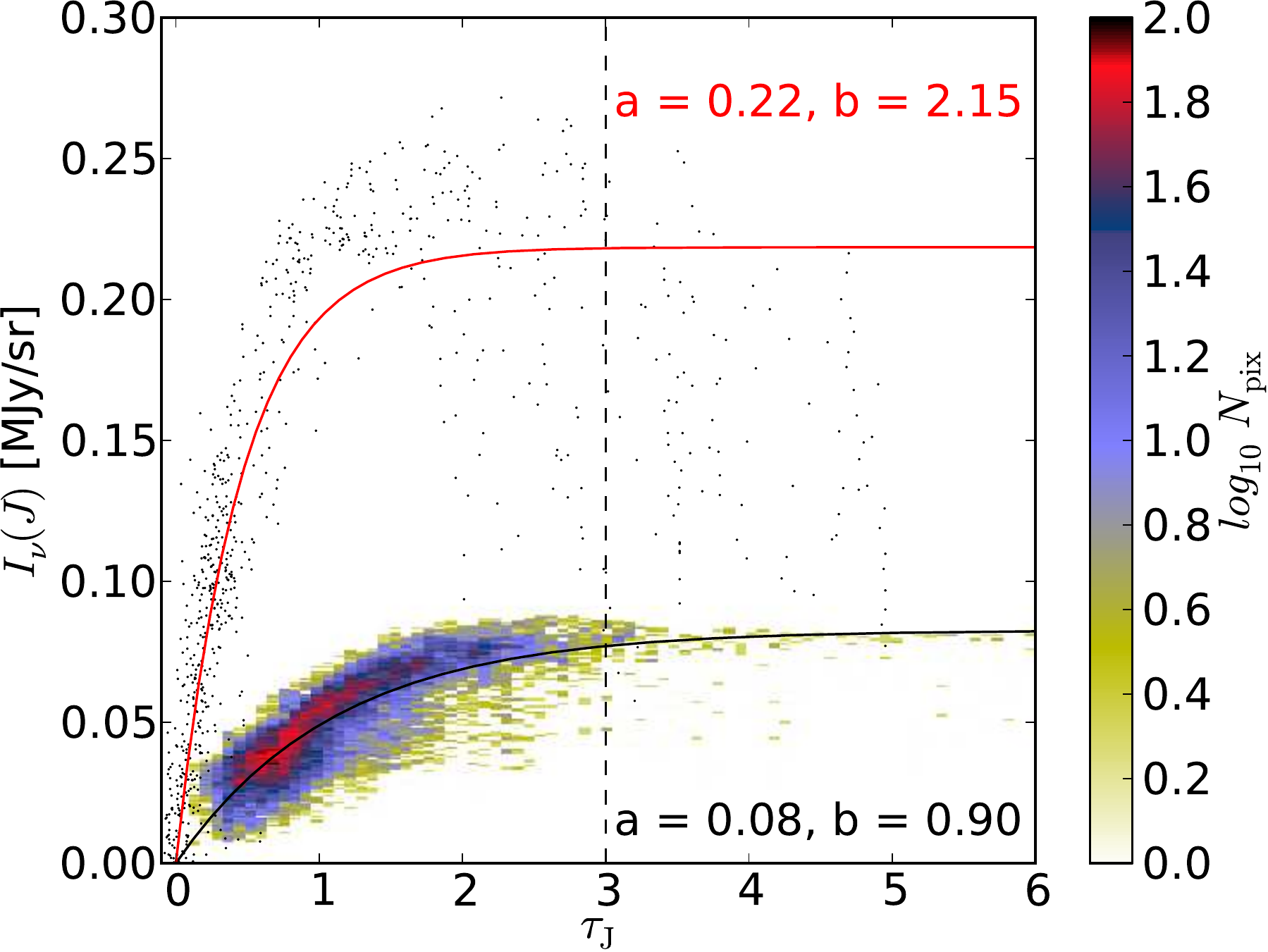}
\includegraphics[width=6cm]{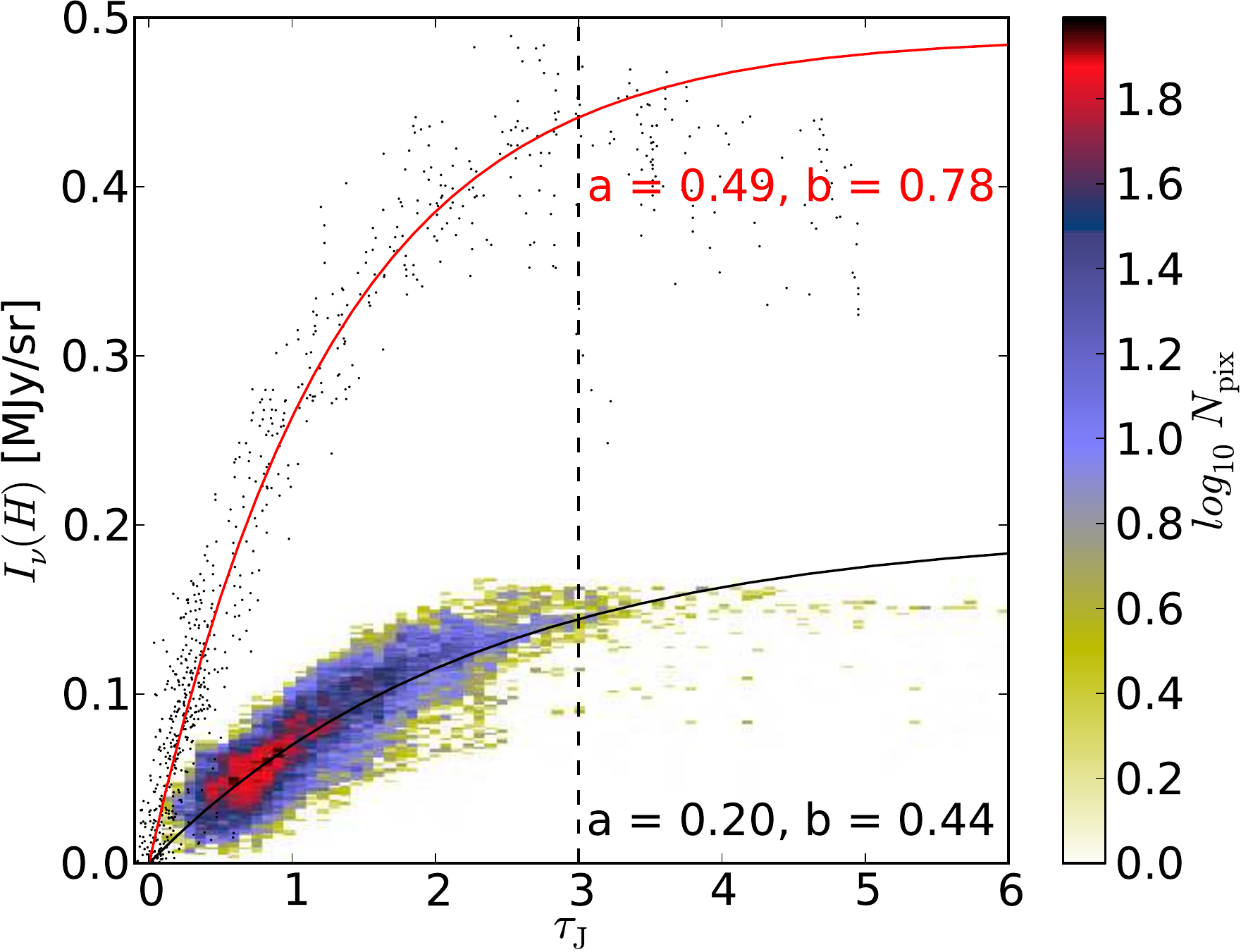}
\includegraphics[width=6cm]{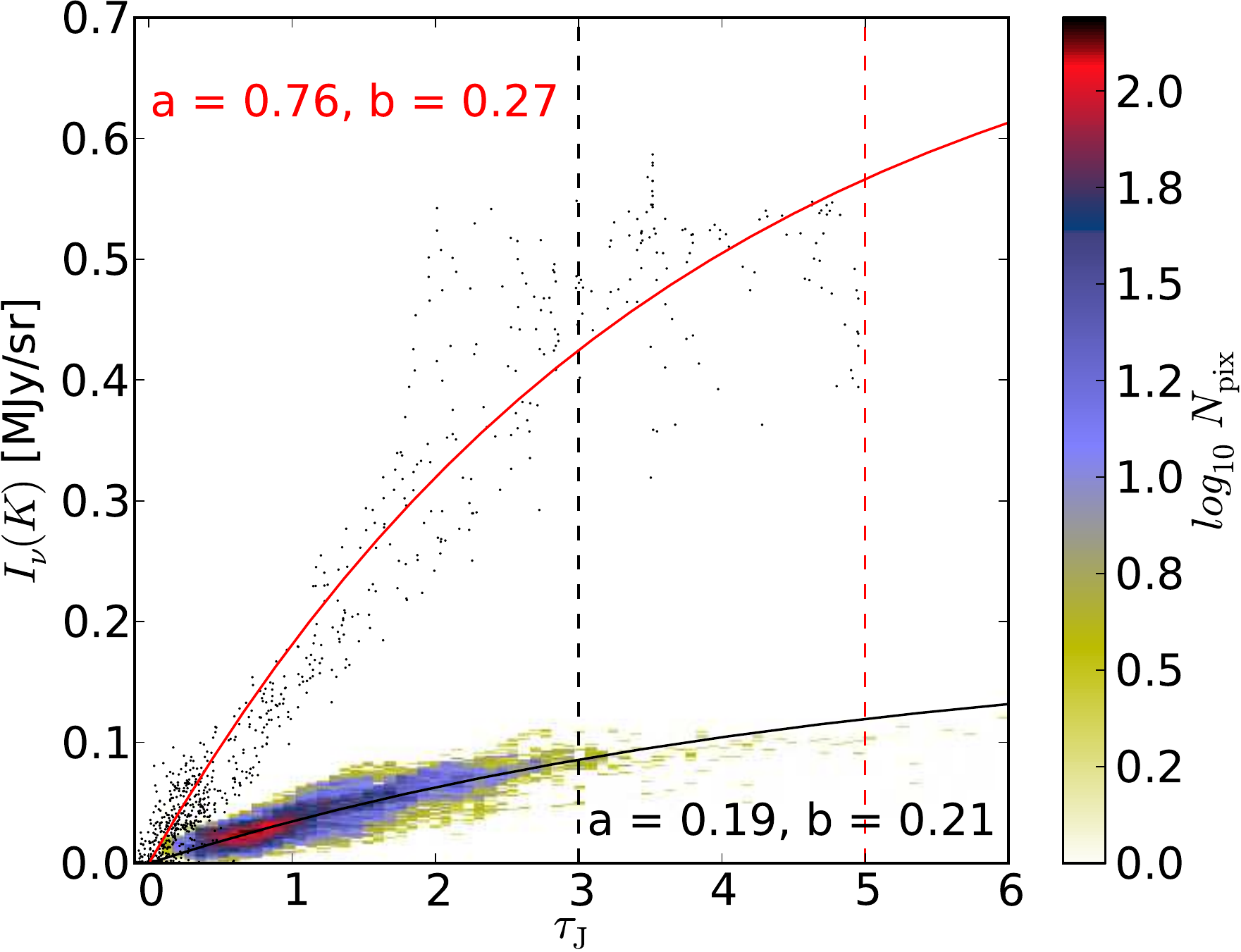}
\caption{Observed NIR surface brightness $I_{\nu}$ as the function of optical depth derived with NICER method $\tau_{J}^{Nicer}$ in $J$, $H$, and $K$ bands. WFCAM data for TMC-1N are shown with a 2D histogram, the colour scale corresponding to the density of points. The black line shows the fitted Eq.~\ref{eq:I_N} (using $\tau_J^{Nicer}$ instead of column density $N$). For comparison, Corona Australis data from~\citet{Juvela2008} are shown with black dots and the fitted function with a red line. The black dashed vertical line shows the upper limit used in the fitting of TMC-1N data in all frames and of Corona Australis data in left and middle frames. In the right frame ($K$ band), the red dashed vertical line shows the upper limit used in the fitting of Corona Australis data, as in this case higher upper limit was needed to make a better fit. The obtained parameter values are marked in the figures with black for TMC-1N and red for Corona Australis.}
\label{fig:scpl_tauJ_I}
\end{figure*}

\begin{figure*}
\centering
\includegraphics[width=6cm]{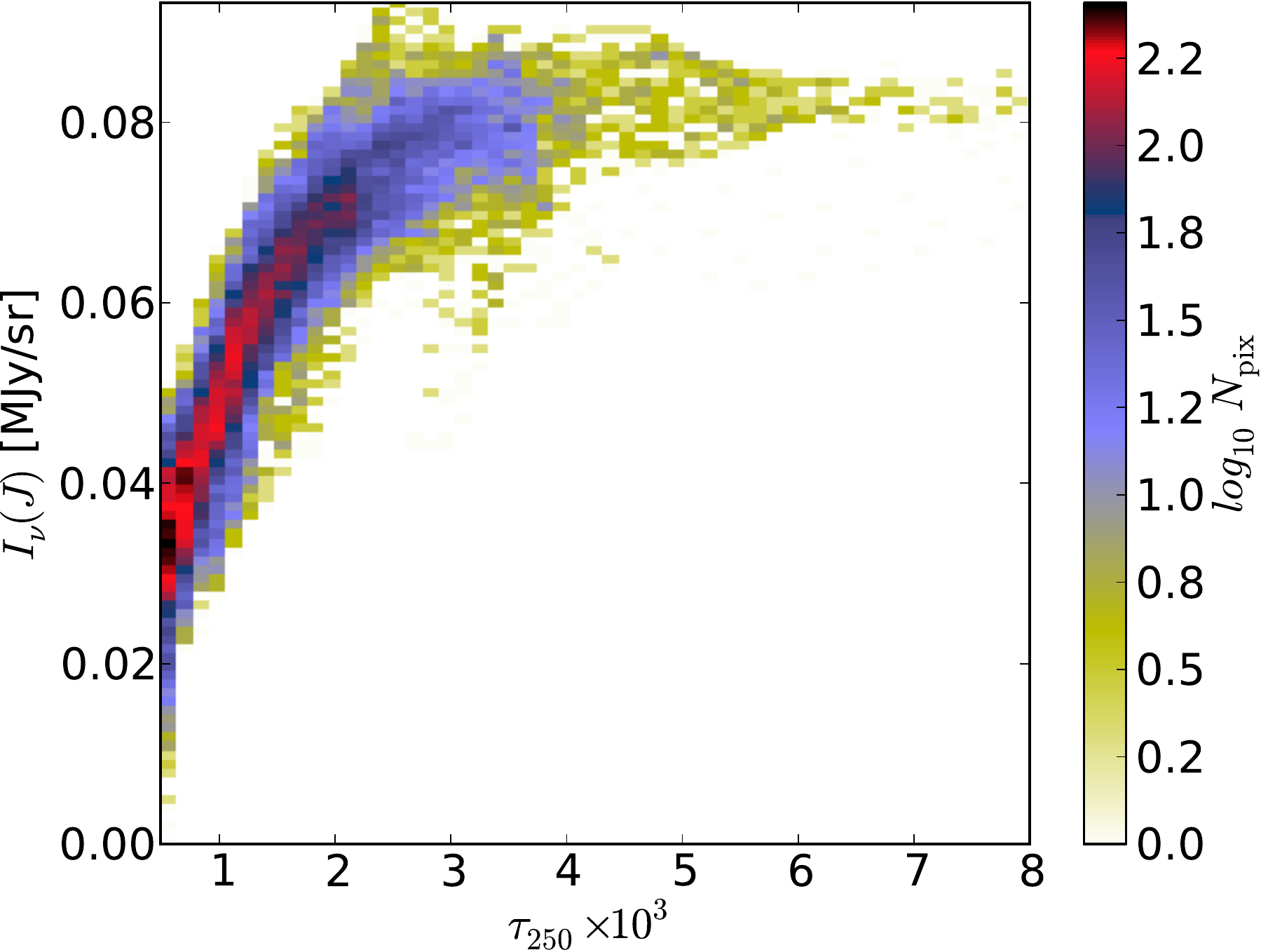}
\includegraphics[width=6cm]{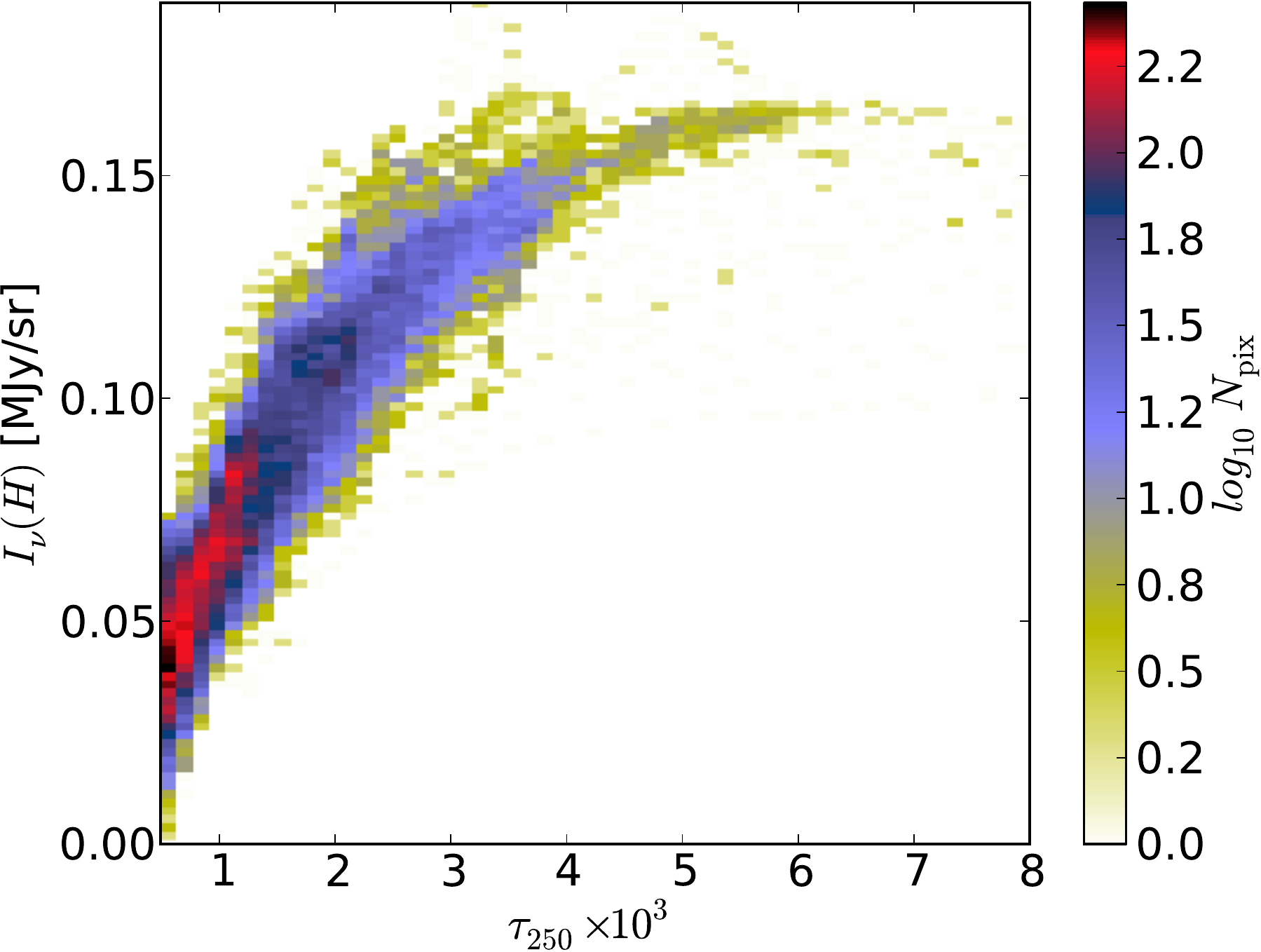}
\includegraphics[width=6cm]{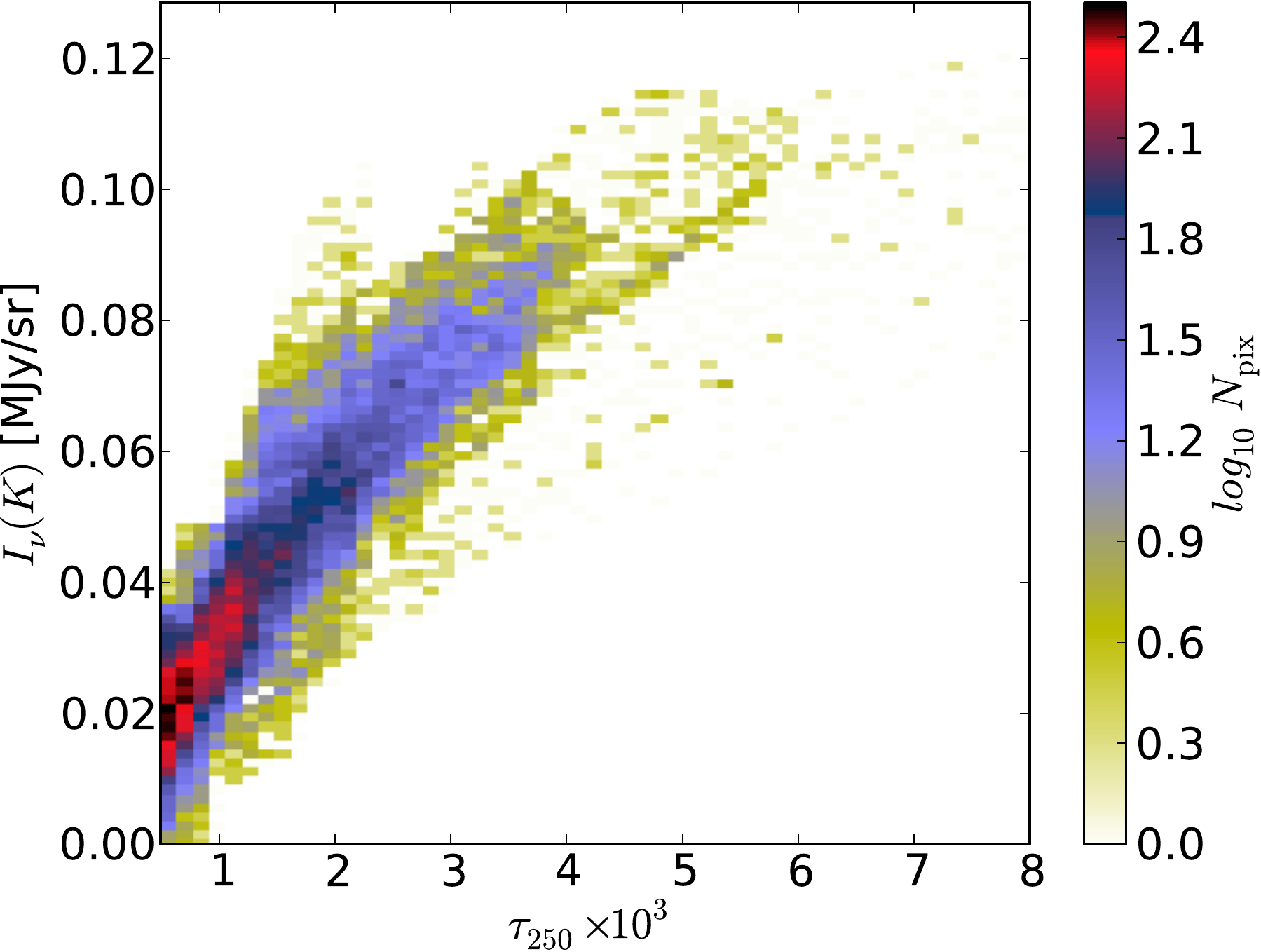}
\caption{Observed NIR surface brightness $I_{\nu}$ as the function of optical depth derived from \emph{Herschel} maps $\tau_{250}$ in $J$, $H$, and $K$ bands. 
The data are at 40" resolution.}
\label{fig:scpl_tau250_I}
\end{figure*}

Correlations between surface brightness, $I_{\nu}$, and optical depth in $J$ band derived from extinction, $\tau_{J}^{Nicer}$, are shown for bands $J$, $H$, and $K$ in Fig.~\ref{fig:scpl_tauJ_I}. We fit Eq.~\ref{eq:I_N} to the data, using optical depth $\tau_J^{Nicer}$ instead of column density $N$ in the exponential part. As a result, we obtain for each band the $a$ and $b$ parameters needed for the derivation of the optical depth map. The fitted values are $a_J=0.08$, $b_J=0.90$, $a_H=0.20$, $b_H=0.44$, $a_K=0.19$, and $b_K=0.21$.
For comparison, we plot in the same figures the data of Corona Australis from~\citet{Juvela2008}. We have scaled those data to $I_{\nu}$ units (MJy/sr), as a function of $\tau_{J}$. Compared to Corona Australis, we obtain notably lower values for $I_{\nu}$ as a function of $\tau_{J}$ in all three bands. We also fit Eq.~\ref{eq:I_N} to these data and show the obtained parameters in the figure.
For comparison, we also show correlations between $I_{\nu}$ and optical depth derived from \emph{Herschel} maps, $\tau_{250}$, in Fig.~\ref{fig:scpl_tau250_I}.

We made error estimates for the $a$ and $b$ parameters obtained from the fitting of Eq.~\ref{eq:I_N} using two different methods. First, we used a standard bootstrap method to estimate the errors caused by sampling. We made the fit using 100 different samples from the data, all samples having the size of the full dataset. The standard deviations for the $a$ and $b$ parameters and their product $a \times b$, that is the slope of the linear part of the function, are shown in Table~\ref{tab:error_estimates}. The standard deviation for parameter $a$ can be up to 0.004, and for parameter $b$ up to $\sim$0.006, but for the slope the uncertainty is rather small, $\sim$0.0002 for $K$, $\sim$0.0003 for $J$, and $\sim$0.0004 for $H$ band.

Secondly, we tested the possible systematic effect of changing the upper limit of the fitted range between $\tau_J$ values $1.5-6.0$. In Table~\ref{tab:error_estimates_XLIMIT} we show the relative change in the parameter values when the upper limit of the fitted range is changed from $\tau_J=2.0$ to $\tau_J=6.0$, as between these values the change is mainly systematic. Even though the change in the $a$ parameter can be up to -33\% and in the $b$ parameter up to 61\%, the change in the slope of the linear part is less than 10\% for each band. Below a $\tau_J$ value of 2.0, the parameter values can change in a more unpredictable way.

\begin{table}
\caption{Standard deviations ($\sigma$) for the $a$ and $b$ parameters and their product $a \times b$.}
\label{tab:error_estimates}
\begin{tabular}{llll}
\hline \hline
Band & $\sigma (a)$ & $\sigma (b)$ & $\sigma (a \times b)$\\
\hline
$J$ & 0.0003 & 0.0062 & 0.00028\\
$H$ & 0.0020 & 0.0060 & 0.00036\\
$K$ & 0.0042 & 0.0056 & 0.00018\\
\hline
\end{tabular}
\end{table}

\begin{table}
\caption{Relative change in the $a$ and $b$ parameters and their product $a \times b$ when the upper limit of the fitted range is changed from $\tau_J=2.0$ to $\tau_J=6.0$.
}
\label{tab:error_estimates_XLIMIT}
\begin{tabular}{llll}
\hline \hline
Band & $\Delta a$ & $\Delta b$ & $\Delta (a \times b)$\\
\hline
$J$ & -0.06 & 0.11 & 0.043\\
$H$ & -0.18 & 0.30 & 0.075\\
$K$ & -0.33 & 0.61 & 0.080\\
\hline
\end{tabular}
\end{table}

\begin{figure}
\centering
\includegraphics[width=9cm]{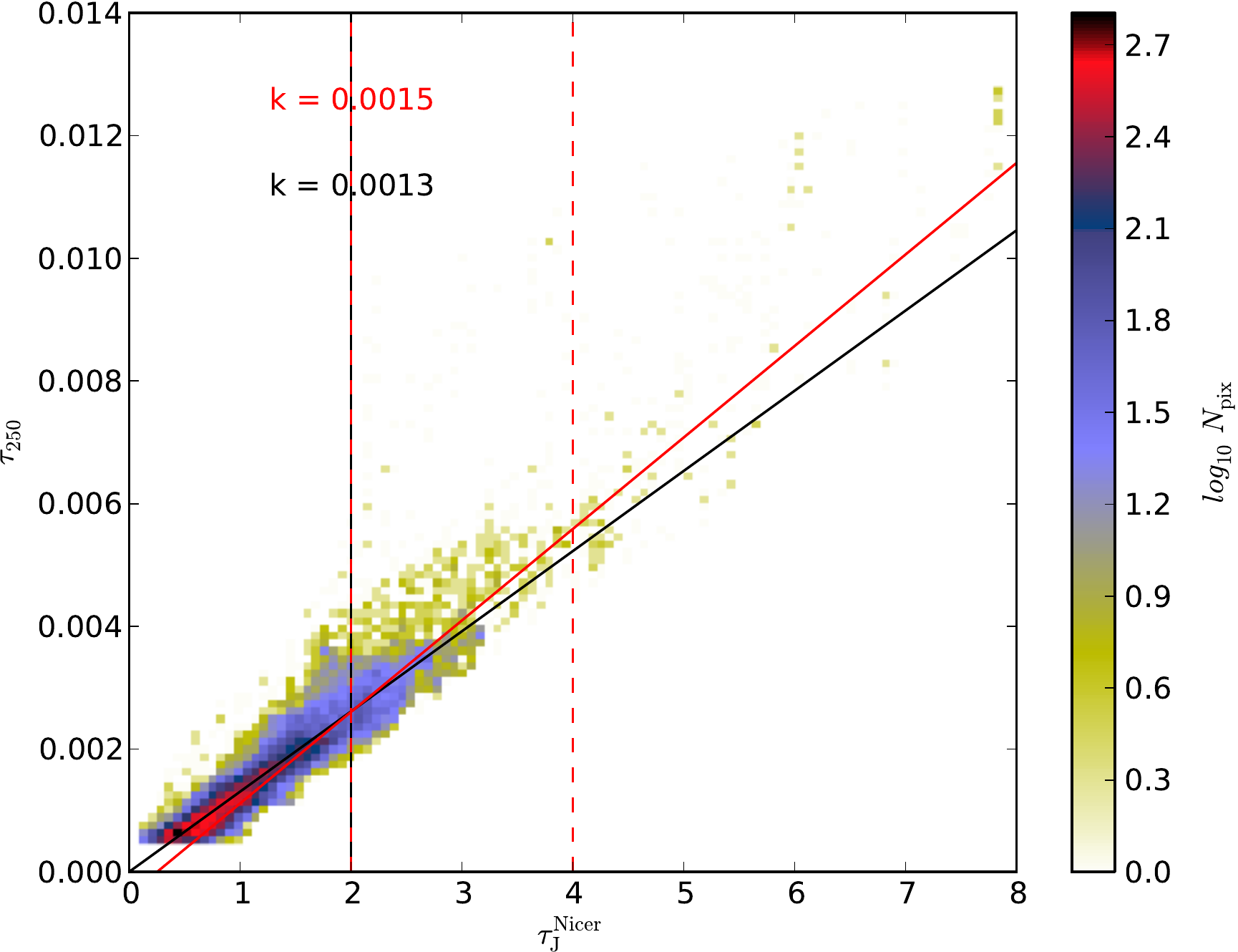}
\caption{Correlation between $\tau_{250}$ and $\tau_{J}^{Nicer}$ with $\beta = 1.8$ for the main filament area shown in Fig.~\ref{fig:I_J_masked}. $\tau_{250}$ map is convolved to the same 60$''$ resolution as $\tau_{J}^{Nicer}$ map. The data are fitted with a straight line, 
using data in the ranges $\tau_J=$0--2 (black line) and $\tau_J=$2--4 (red line). The values of the slopes, $k$, are indicated in the frame with corresponding colours.}
\label{fig:scpl_tauNJ_tau250}
\end{figure}

We derive emissivity, the ratio of FIR dust emission to column density, by using $\tau_{J}$ obtained from NIR extinction map as an independent tracer of column density. Correlation between $\tau_{250}$ and $\tau_{J}^{Nicer}$ with $\beta = 1.8$ is shown in Fig.~\ref{fig:scpl_tauNJ_tau250} for the main filament area shown in Fig.~\ref{fig:I_J_masked}. The slope of a straight line fitted to the range $\tau_{J}^{Nicer}=$0--4 is $\sim$0.0013. 
We also tested the possible change of slope in the low and middle $\tau_{J}^{Nicer}$ range. The derived slope is 0.0013 for range 0--2, and 0.0015 (more precisely 0.00149) for range 2--4.
If we change $\beta$ from 1.8 to 2.0, the slope between 0--4 increases $\sim$32\% to the value $\sim$0.0018.
We also made a similar fit using all the data in the maps. The derived slope is 0.0013 for range 0--2, and 0.0014 (more precisely 0.00144) for range 2--4, indicating that there is no significant change when compared to the masked area.

For comparison with other studies, we convert our slope of $\tau_{250}/\tau_{J} = 0.0013$ to opacity or dust emission cross-section per H nucleon
\begin{equation}
\sigma_e(\nu) = \tau_{\nu}/N_{\rm H} = \mu m_{\rm H} \kappa_{\nu} [{\rm cm}^2/{\rm H}],
\label{eq:sigma}
\end{equation}
where $\tau_{\nu}$ is optical depth, $N_{\rm H}$ is the total H column density (H in any form), $\mu$ is the mean molecular weight per H (1.4), $m_{\rm H}$ is the mass of H atom, and $\kappa_{\nu}$ is the mass absorption (or emission) coefficient (cm$^2$/g) relative to gas mass, also often called opacity. We again use wavelength instead of frequency in our notation: $\sigma_e(250) = \sigma_e(250\mu {\rm m}) = \sigma_e(1200{\rm GHz})$.

We derive the $\tau_{250}$ map directly from the \emph{Herschel} observations, and use our WFCAM NIR extinction map as an independent tracer of the column density.
We use the conversion factor for diffuse clouds
$N({\rm HI} + {\rm H}_2)/E(B-V) = 5.8 \times 10^{21}$ cm$^{-2}$/mag
of \citet{Bohlin1978}. We derive the relation $E(B-V)/E(J-K)$ from \citet{Cardelli1989} extinction curves, and obtain values 1.999 (with $R_V = 3.1$) and 1.413 (with $R_V = 4.0$), leading to relations $N({\rm H}) = 11.59 \times 10^{21} E(J-K)$ (with $R_V = 3.1$) and $N({\rm H}) = 8.196 \times 10^{21} E(J-K)$ (with $R_V = 4.0$). \citet{Martin2012} have observed a similar relation, $N({\rm H}) \sim 11.5 \times 10^{21} E(J-K_S)$, for regions of moderate extinction in Vela.
\citet{Cardelli1989} extinction curves also give the relation
$A_J/E(J-K) \sim 1.675$ (with both $R_V$ values 3.1 and 4.0). Converting magnitudes to optical depths ($A = 2.5 lg(e)\tau \sim 1.086\tau$) gives the relation $E(J-K) \sim \tau_J/1.54$. This leads to the relations $\sigma_e(250) = 1.33 \times 10^{-22} \tau_{250}/\tau_{J}$ cm$^2$/H (with $R_V = 3.1$) and $\sigma_e(250) = 1.88 \times 10^{-22} \tau_{250}/\tau_{J}$ cm$^2$/H (with $R_V = 4.0$).

The value $\tau_{250}/\tau_{J} = 0.0013$ leads to values $\sigma_e(250) = 1.7 \times 10^{-25} {\rm cm}^2/{\rm H}$ (with $R_V = $3.1) or $\sigma_e(250) = 2.4 \times 10^{-25} {\rm cm}^2/{\rm H}$ (with $R_V = 4.0$). These can be converted to $\kappa_{\nu}$ values 0.07 cm$^2$/g or 0.10 cm$^2$/g, respectively.

\begin{figure}
\centering
\includegraphics[width=9cm]{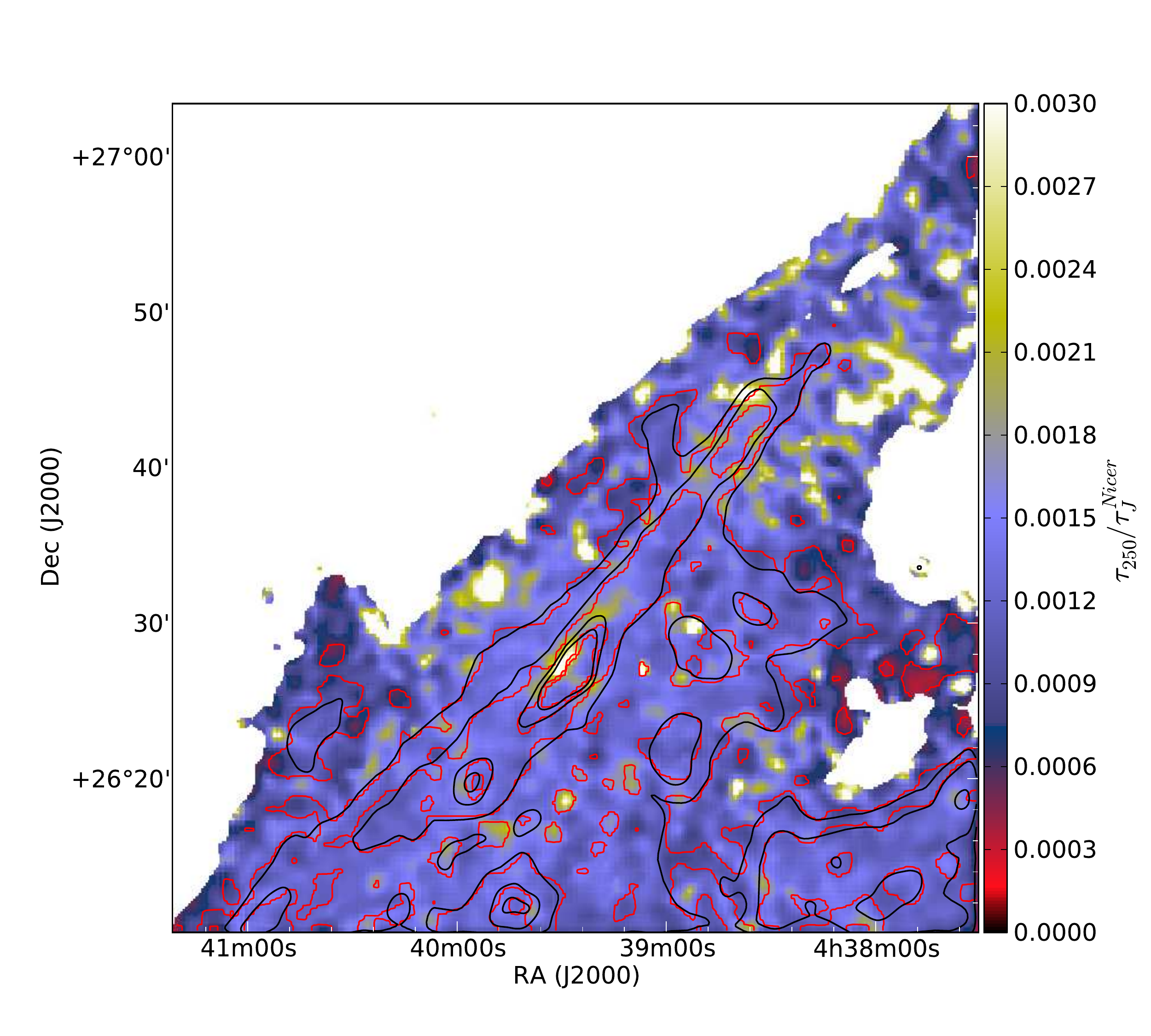}
\caption{$\tau_{250}/\tau_{J}^{Nicer}$ map. $\tau_{250}$ is derived from \emph{Herschel} maps. $\tau_{250}$ map is convolved to the same 60$''$ resolution as $\tau_{J}^{Nicer}$ map. Areas where $\tau_{250} < 0.0003$ are masked. Contours for $\tau_{J}^{Nicer}=$ 4, 3, 1.5, and 0.75 are marked with red line. Contours for $\tau_{250}=$ 0.0052, 0.0039, 0.0019, and 0.0010 are marked with black line.}
\label{fig:map_tau250_I}
\end{figure}

The map of the ratio $\tau_{250}/\tau_{J}^{Nicer}$ is shown in Fig.~\ref{fig:map_tau250_I}. The map shows no clear evidence for systematic increase of the ratio $\tau_{250}/\tau_{J}$ with increasing density inside the densest filament. Some high value areas can be attributed to imperfections in the two maps, $\tau_{250}$ and $\tau_{J}$. For instance, the high value area next to the densest part of the filament seems to be caused by the different shape of the filament in these two maps. There the value of the ratio is high, because the filament is slightly more narrow in $\tau_{J}$, possibly caused by the lack of background stars seen behind the densest filament.

\subsection{Optical depth derived from scattered light} \label{sect:columndensity}

As described in Sect.~\ref{sect:method}, in order to derive an optical depth $\tau_{J}^{SB}$ map from the surface brightness maps, we obtain $a$ and $b$ parameter values for each band from correlations between $I_{\nu}$ and $\tau_{J}^{Nicer}$ shown in Fig.~\ref{fig:scpl_tauJ_I}. We calculate a value for each pixel of the $\tau_{J}^{SB}$ map by minimising Eq.~\ref{eq:tauSB_1}.

\begin{figure*}
\centering
\includegraphics[width=6cm]{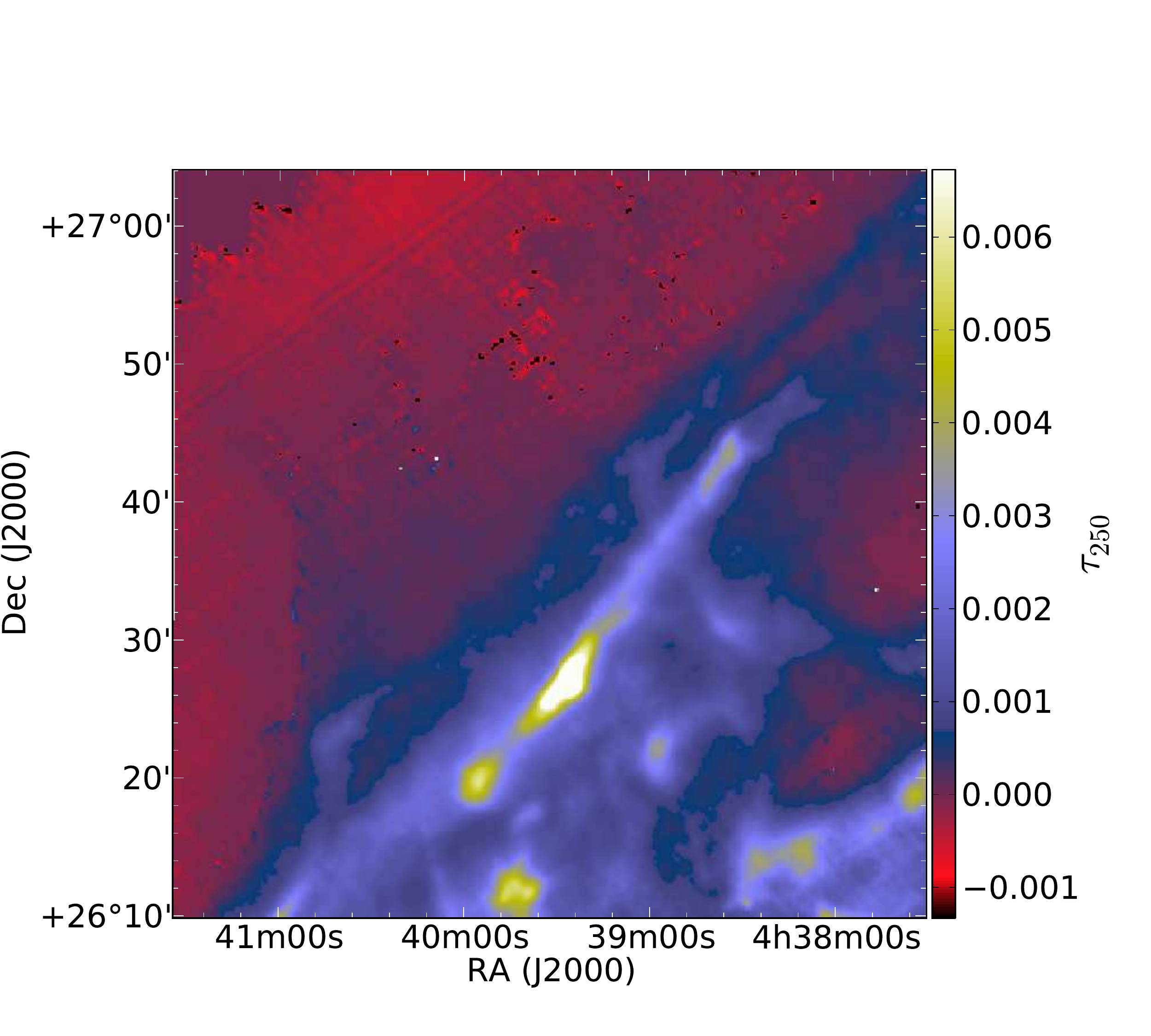}
\includegraphics[width=6cm]{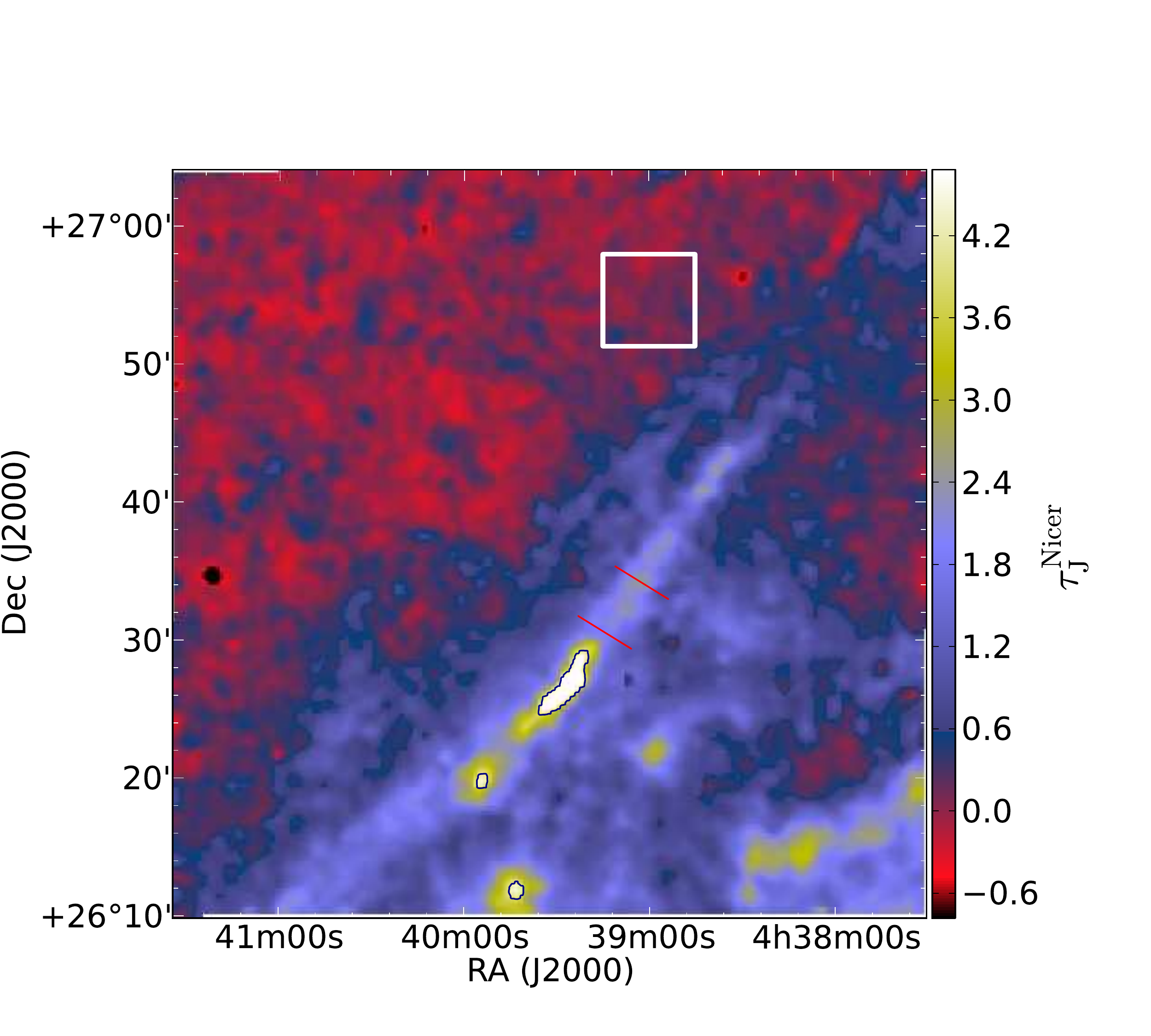}
\includegraphics[width=6cm]{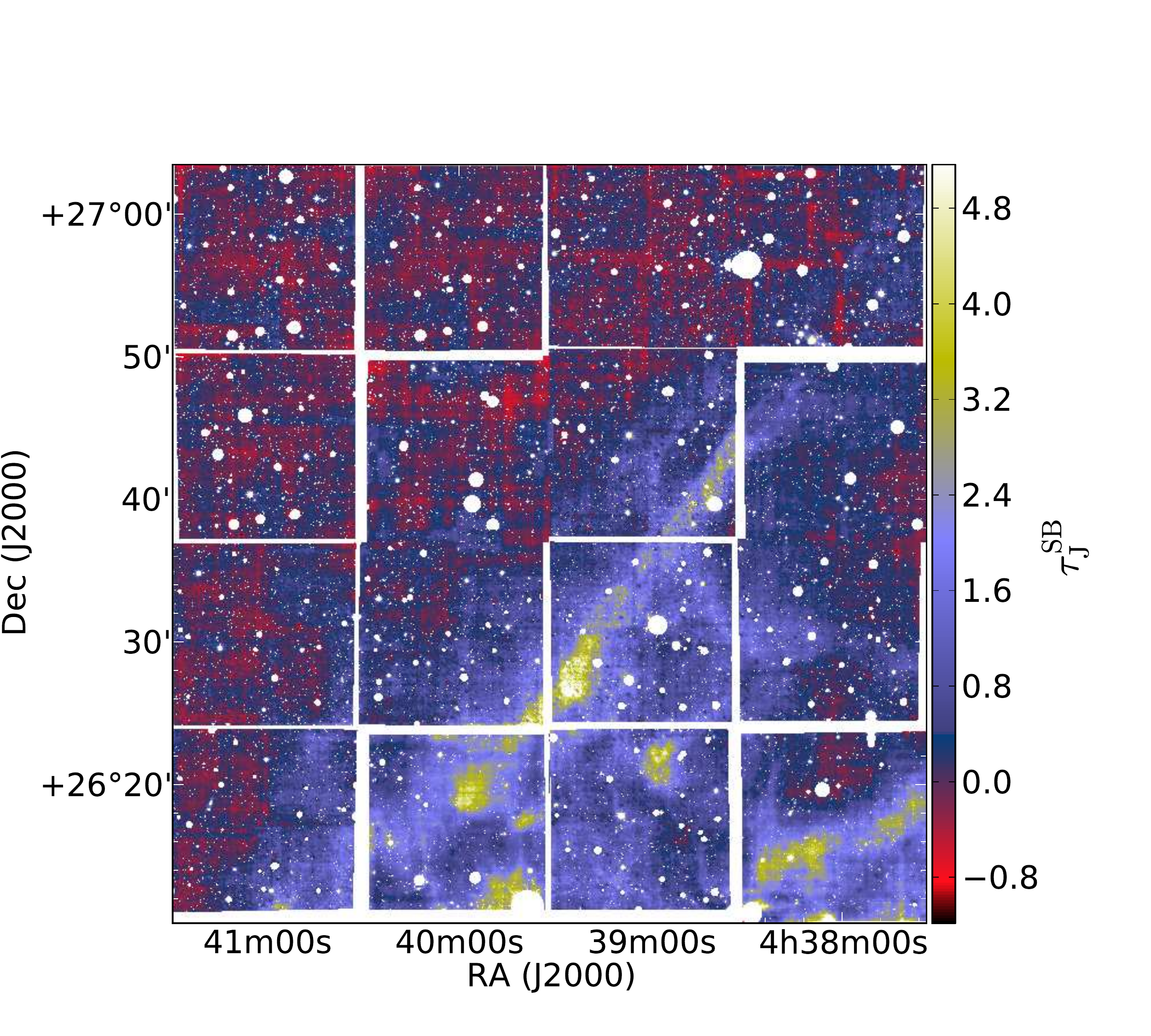}
\caption{Maps of optical depth $\tau$: \emph{Herschel} $\tau_{250}$ (resolution 40$''$), NICER $\tau_{J}^{Nicer}$ (resolution 60$''$), and optical depth in $J$ band, $\tau_{J}^{SB}$, based on $J$, $H$, and $K$ surface brightness maps (resolution $\sim$2.2$''$). In the NICER map (middle frame) the densest areas where $\tau_{J}^{Nicer} > 4$ are marked with a blue contour. The white rectangle marks the reference area used for background subtraction in the analysis. The profile cross-sections shown in Fig.~\ref{fig:median_profile_new} are taken as a median from the area between the two red lines.}
\label{fig:tau_fullmaps}
\end{figure*}

\begin{figure*}
\centering
\includegraphics[width=6cm]{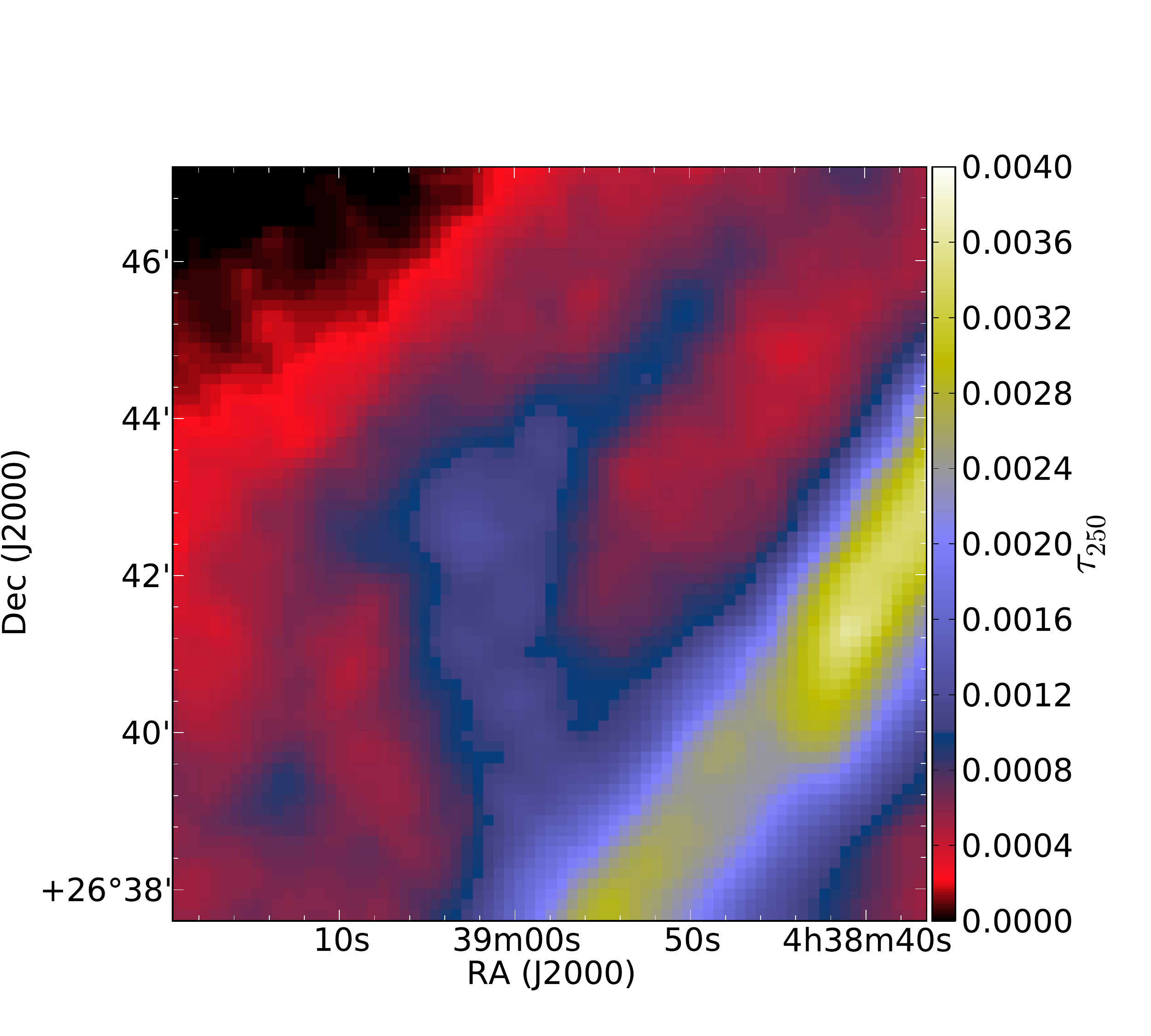}
\includegraphics[width=6cm]{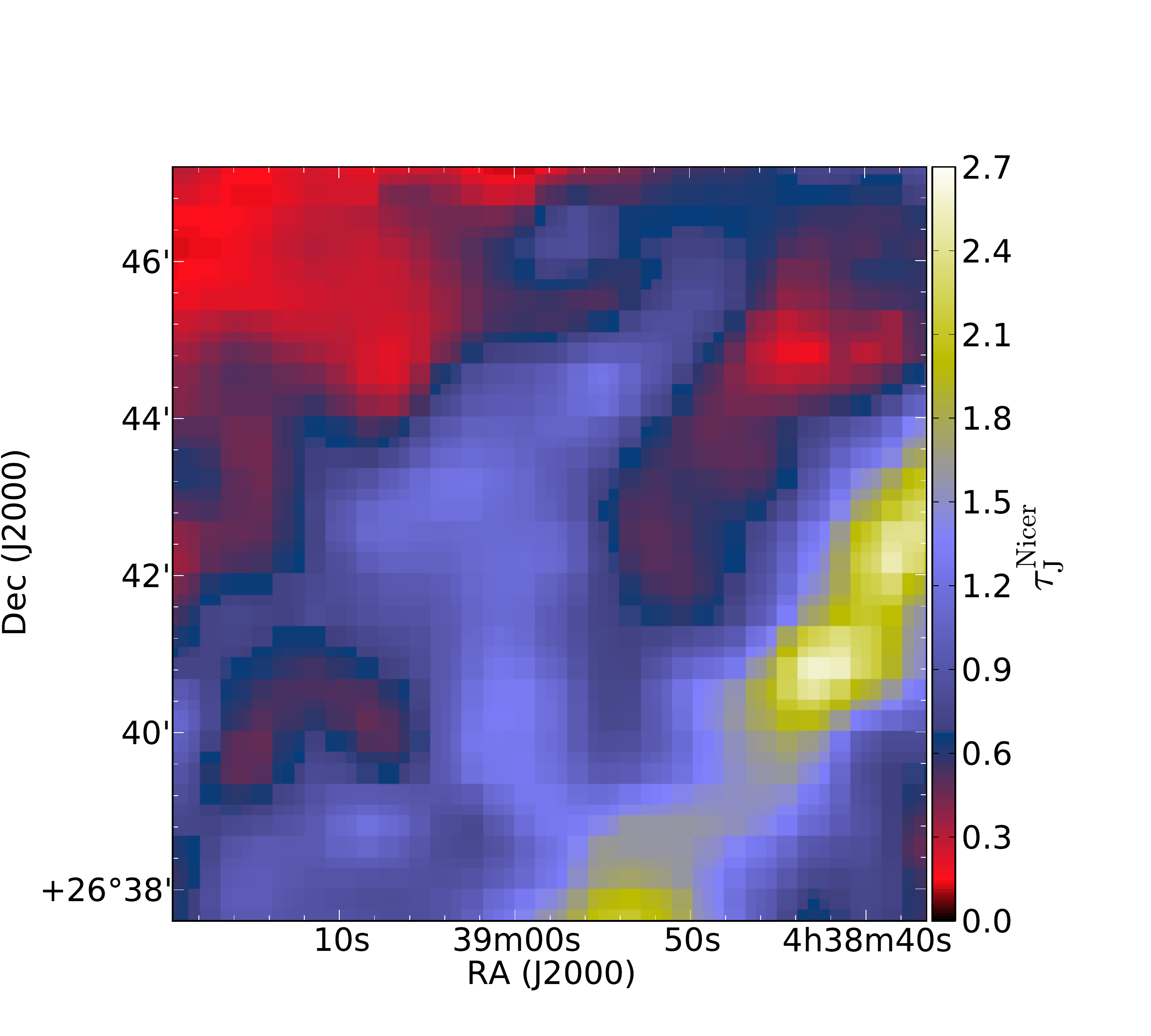}
\includegraphics[width=6cm]{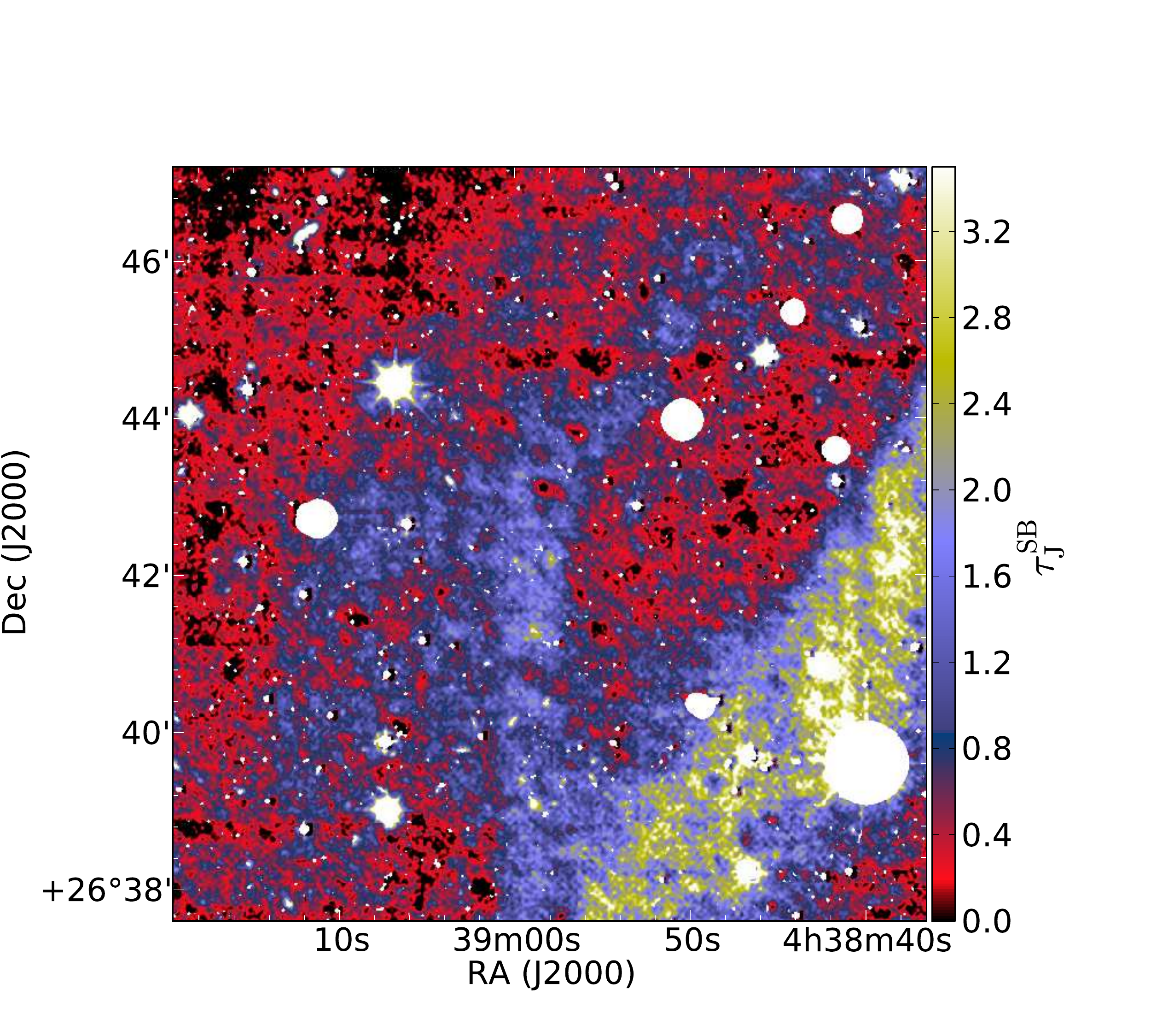}
\caption{Maps of optical depth $\tau$ in a 10$\times$10 arcmin area centered at 4h38m58s, +26$^{\circ}$42$'$24$''$ (RA~(J2000), Dec~(J2000)): \emph{Herschel} $\tau_{250}$ (resolution 40$''$), NICER $\tau_{J}^{Nicer}$ (resolution 60$''$), and $\tau_{J}^{SB}$ (resolution $\sim$2.2$''$).} 
\label{fig:tau_zoomins}
\end{figure*}

We use 40$''$ and 60$''$ maps in the following analysis, but also show a higher resolution ($\sim$2.2$''$) $\tau_{J}^{SB}$ map in Fig.~\ref{fig:tau_fullmaps} (right frame). In the same figure, we also show maps of \emph{Herschel} $\tau_{250}$ and NICER $\tau_{J}^{Nicer}$. In Fig.~\ref{fig:tau_zoomins}, we show close-ups of the same maps.

\begin{figure}
\centering
\includegraphics[width=9cm]{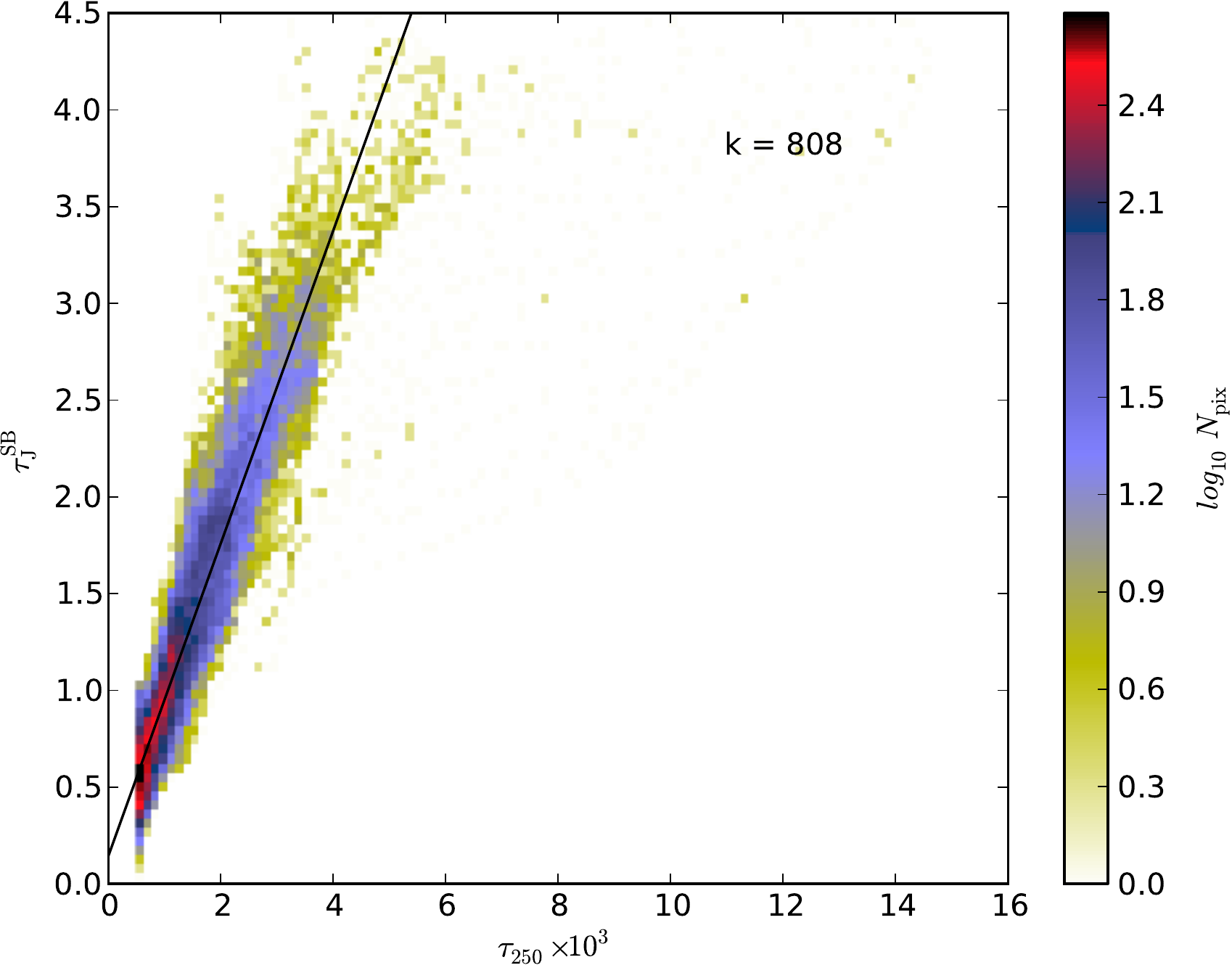}
\includegraphics[width=9cm]{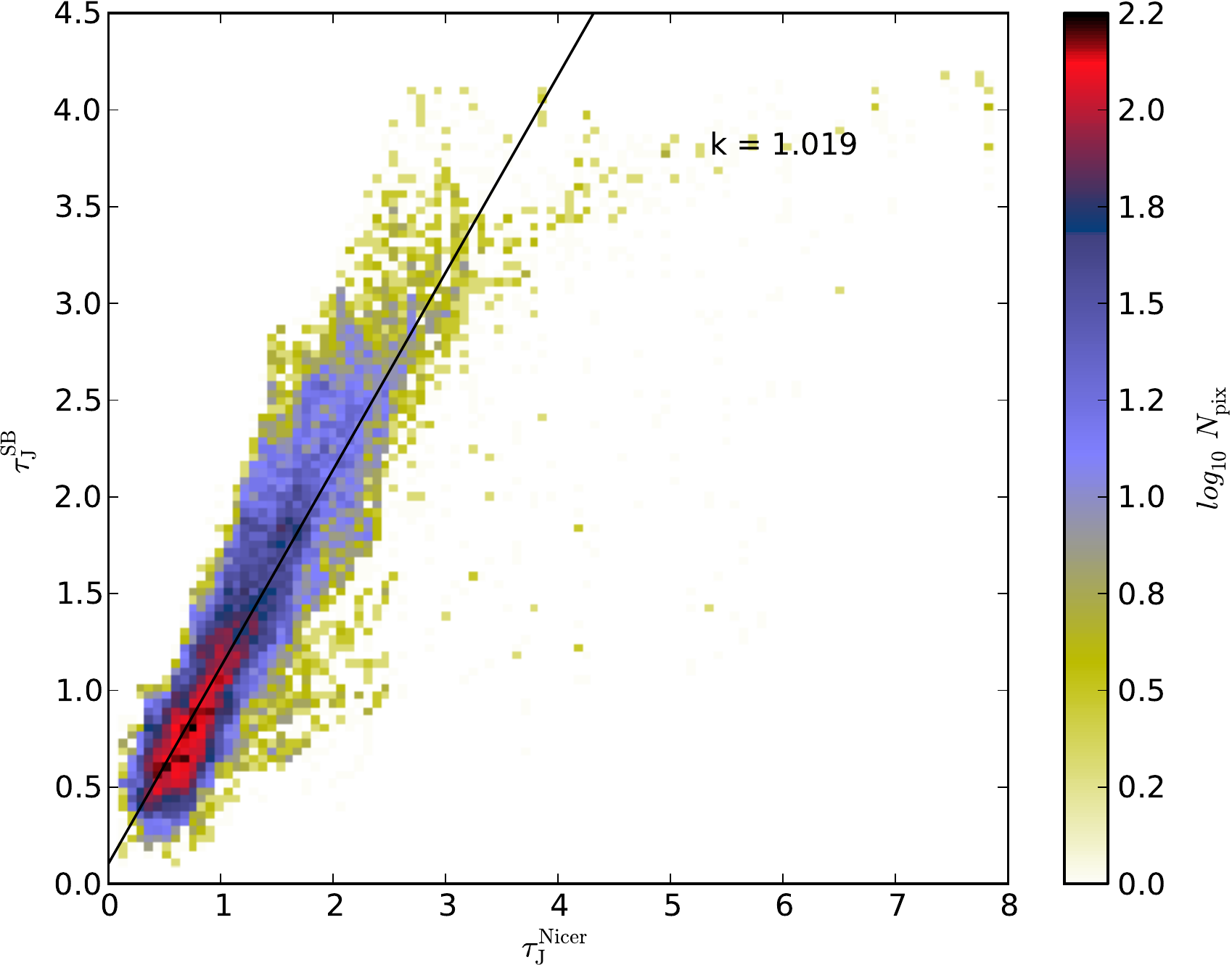}
\caption{Correlations between $\tau_{SB}$ and $\tau_{250}$ with resolution 40$''$ (upper frame) and between $\tau_{SB}$ and $\tau_{J}^{Nicer}$ with resolution 60$''$ (lower frame). The data are fitted with a straight line, marked with black line. The fitted values for the slope, $k$, are marked to the figures.}
\label{fig:plots_SB}
\end{figure}

Correlations between $\tau_J^{SB}/\tau_{250}$ (resolution 40$''$) and $\tau_J^{SB}/\tau_{J}^{Nicer}$ (resolution 60$''$) are shown in Fig.~\ref{fig:plots_SB}. The reference area used for background subtraction is shown in Fig.~\ref{fig:tau_fullmaps} (middle frame). The fitted values for the slopes are 808 for $\tau_J^{SB}/\tau_{250}$ and 1.019 $\sim$ 1 for $\tau_J^{SB}/\tau_{J}^{Nicer}$. This is as expected, since $\tau_J^{SB}$ was derived based on the correlation between $I_{\nu}$ and $\tau_{J}^{Nicer}$. The correlations are linear up to $\tau_{J}^{Nicer} \sim 4$. Above that, the $\tau_{J}^{SB}$ values saturate strongly. The areas where  $\tau_{J}^{Nicer} > 4$ are marked with contours in Fig.~\ref{fig:tau_fullmaps} (middle frame), indicating that these form only a small area in the densest clumps inside the cloud. 

\subsection{Filament cross-sections} \label{sect:filament}

\begin{figure}
\centering
\includegraphics[width=9cm]{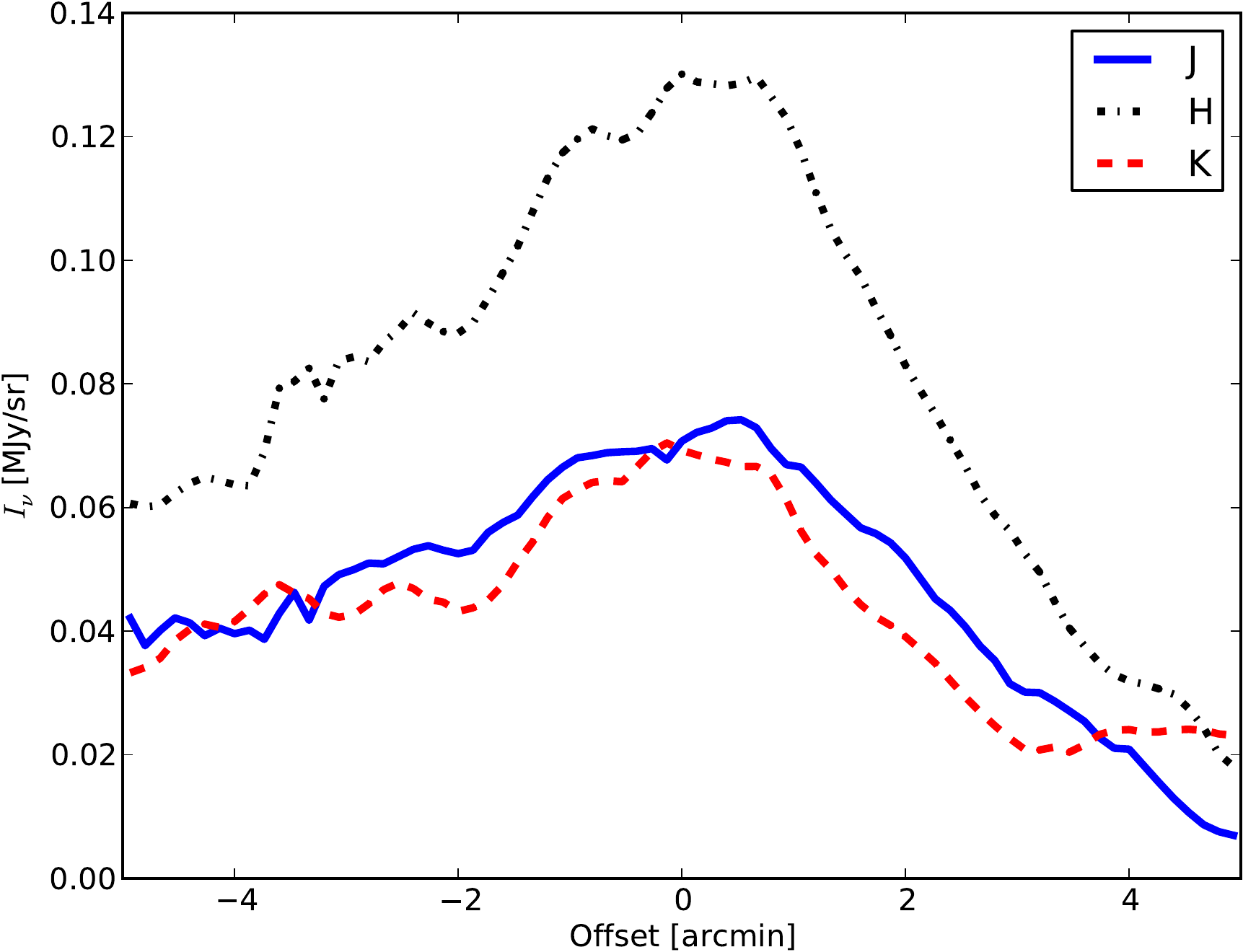}
\includegraphics[width=9cm]{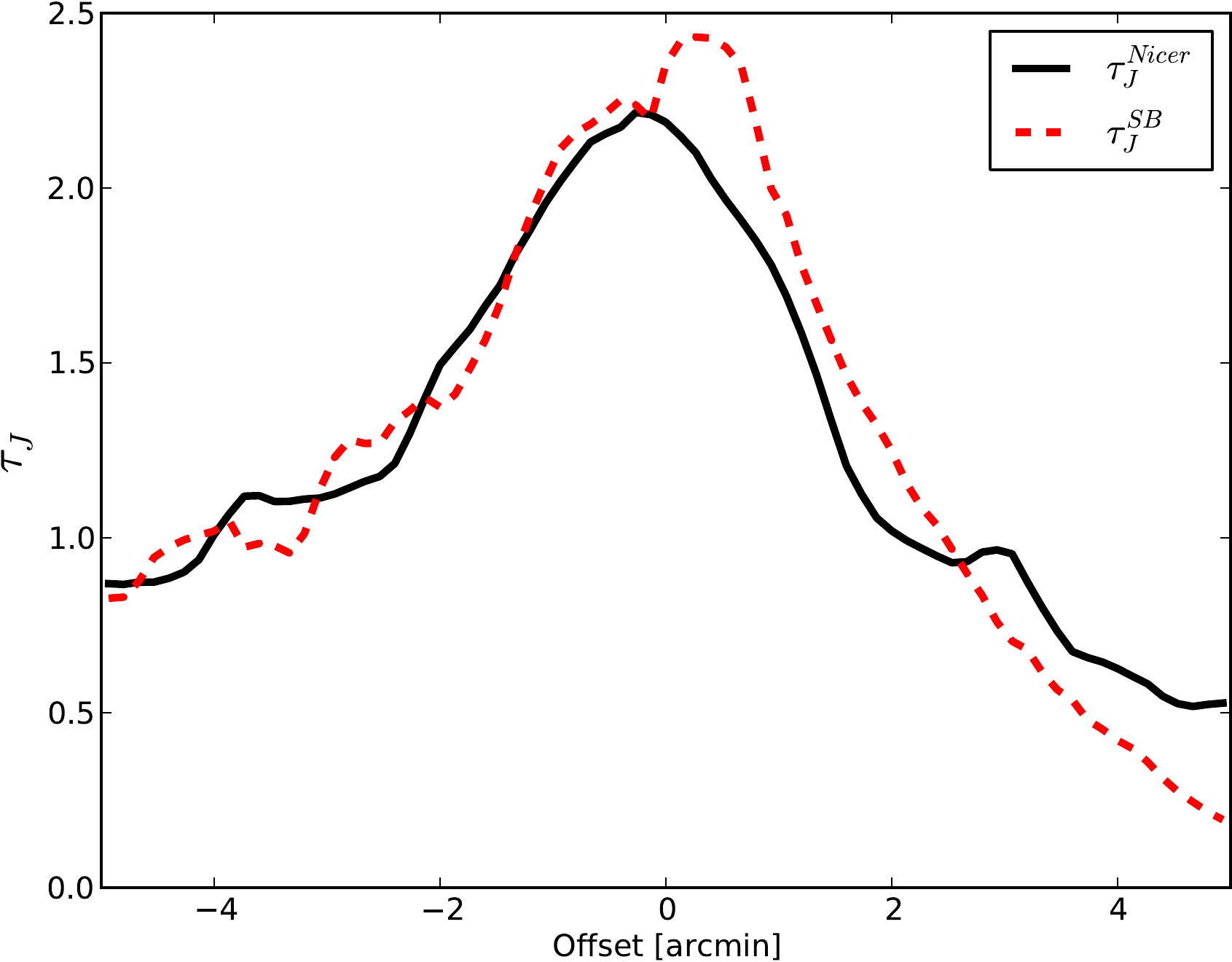}
\caption{Median profile cross-sections of the filament, taken from the area between the two red lines shown in Fig.~\ref{fig:tau_fullmaps} (middle frame). (Upper frame) WFCAM maps in $J$, $H$, and $K$ bands. (Lower frame)  $\tau_{J}^{Nicer}$ is derived from NICER extinction map and $\tau_{J}^{SB}$ from WFCAM surface brightness maps, respectively.}
\label{fig:median_profile_new}
\end{figure}

Median profiles of the filament are shown in Fig.~\ref{fig:median_profile_new}, for bands $J$, $H$, and $K$ (upper frame) and for $\tau_{J}^{Nicer}$ and $\tau_{J}^{SB}$ (lower frame). The cross-sections are taken from the area between the two red lines shown in Fig.~\ref{fig:tau_fullmaps} (middle frame), where we have continuous data in all WFCAM maps. This is not the densest part of the filament, but a moderately dense area next to it. The cross-sections show that the derived $\tau_{J}^{SB}$ gives rather similar results to the $\tau_{J}^{Nicer}$ map for the filament profile, except for an extra peak in the $\tau_{J}^{SB}$ profile. The profiles seen in $J$ and $K$ bands are rather similar, whereas the $H$ band profile has approximately two times stronger peak than the other two bands. The filament width or FWHM is $\sim3' \sim 0.1$ pc, as shown in~\citet{Malinen2012}.

\subsection{Spitzer data} \label{sect:MIR-data}

We compare the IRAC 3.6\,$\mu$m surface brightness with the \emph{Herschel} optical depth $\tau_{250}$, to study also the correlations between MIR scattered light and sub-millimetre dust emission. The surface brightness at $\sim$3.6\,$\mu$m can still be dominated by light scattering, while at longer wavelengths, the scattering decreases and the signal is expected to be dominated by dust emission. Some contribution of dust emission cannot be excluded even around 3.6\,$\mu$m and we conservatively consider the Spitzer data only as an upper limit on the intensity of the scattered light.

The point sources are a major problem in estimating the level of the extended emission.
Because the examined area is small, we can produce an extended MIR surface brightness map by manually masking all the visible point sources.
The masks altogether cover 24\,arcmin$^2$.
After the masking, we carry out median filtering. The filter calculates the 25\% percentile of all unmasked pixels within a given radius that is fixed to 5$''$. If the area contains less than ten unmasked pixels, the value is left undefined.
The images are then convolved with a Gaussian beam to produce final surface brightness maps at the resolution of 40$''$ that corresponds to the resolution of the \emph{Herschel} data. The convolution ignores the undefined values.

To reduce the noise (and because of the uncertainty of the fidelity of the large scale flat-fielding), data are correlated only around the main column density peak. The column density threshold of 4$\times 10^{21}$\,cm$^{-2}$ defines an area of $\sim$77\,arcmin$^2$.

\begin{figure}
\centering
\includegraphics[width=9cm]{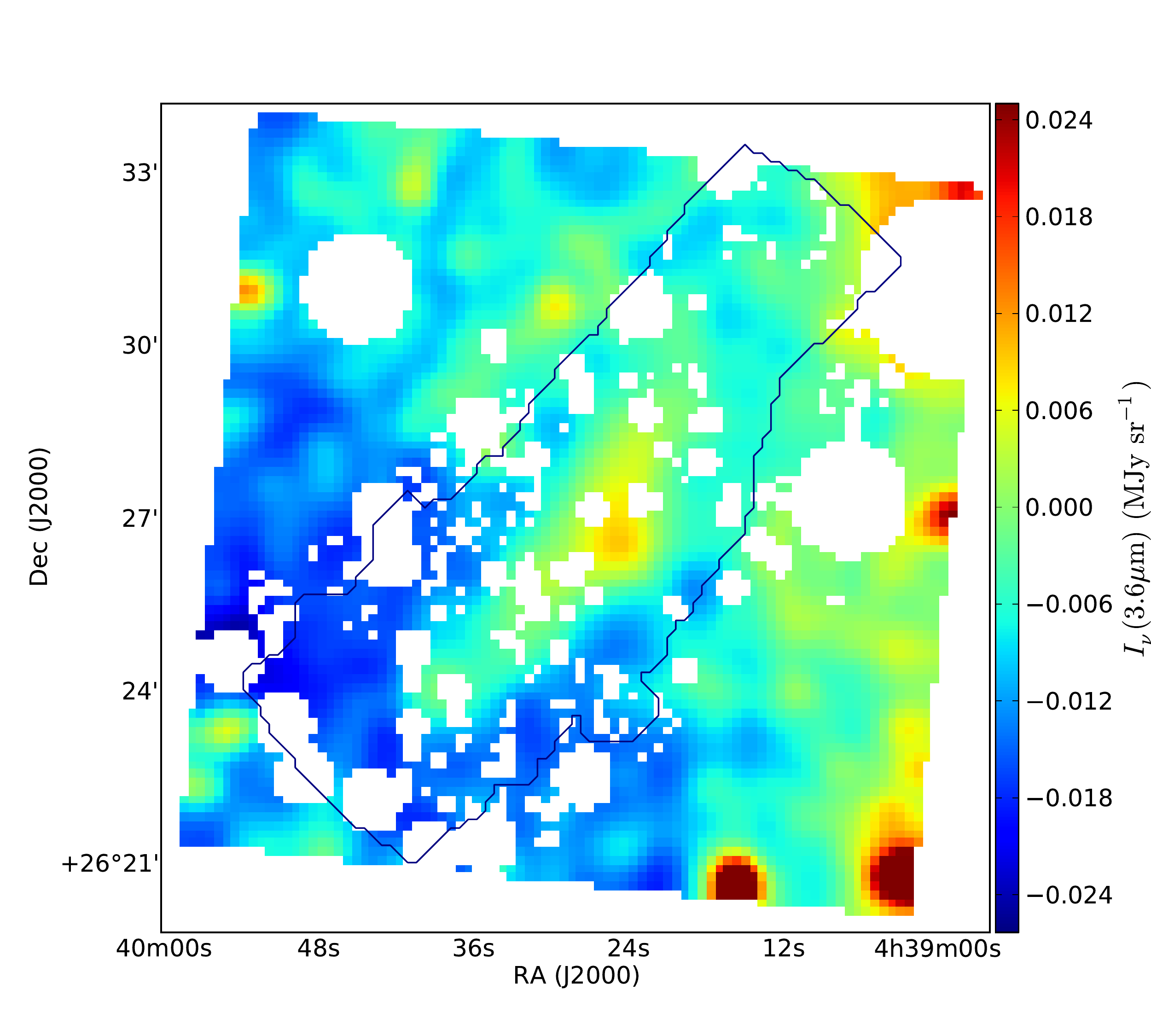}
\caption{
Spitzer 3.6\,$\mu$m intensity map of TMC-1N. The area with highest column density is shown with a contour and used in the scatter plot in Fig.~\ref{fig:Sp_plot}. Pixels affected by point sources are masked and left empty in the figure.
}
\label{fig:Sp_map}
\end{figure}

\begin{figure}
\centering
\includegraphics[width=9cm]{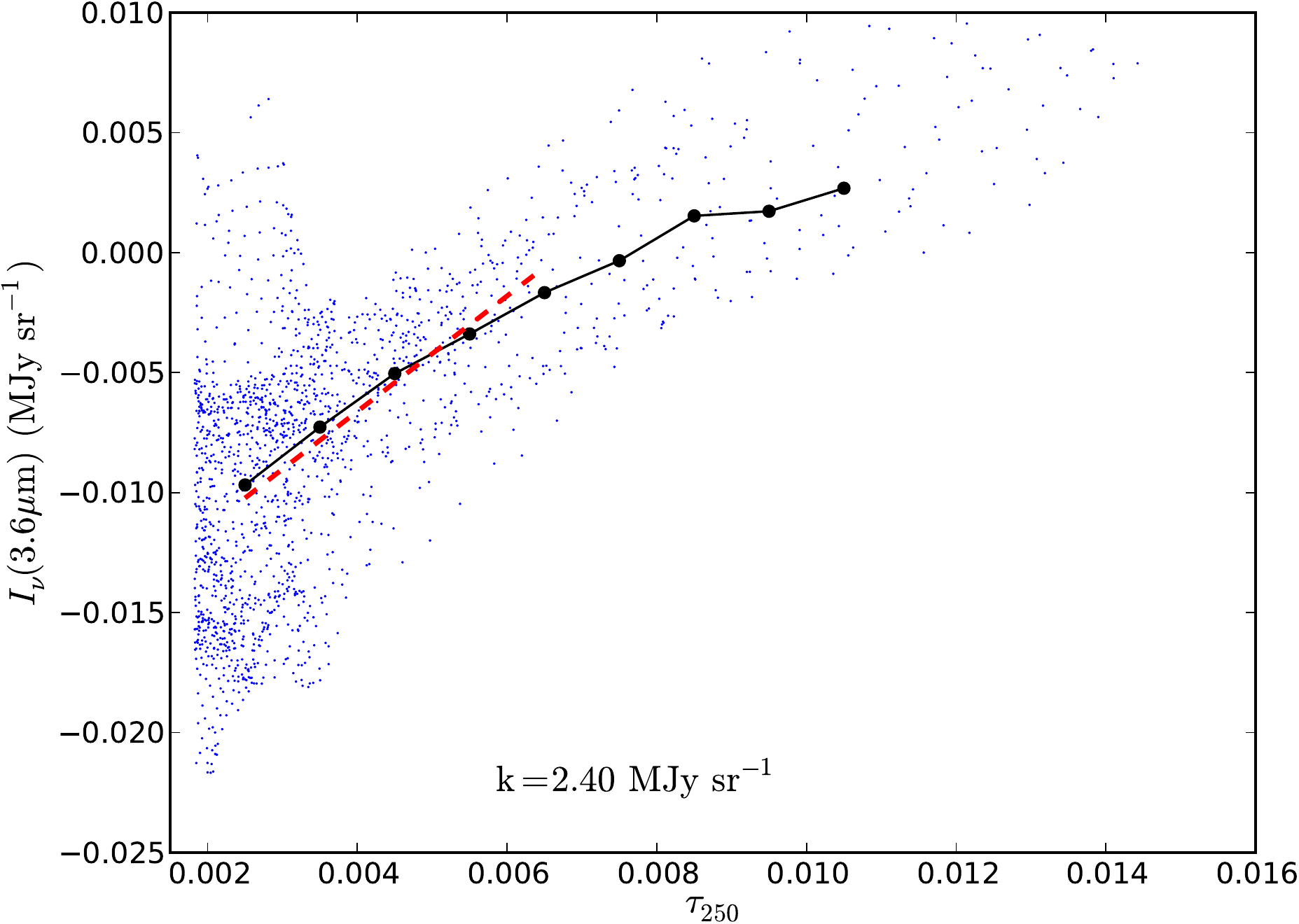}
\caption{
Spitzer 3.6\,$\mu$m surface brightness as the function of $\tau_{250}$ derived from \emph{Herschel} data. The black circles show the median surface brightness in $\tau_{250} = $ 0.002 wide bands. The dashed red line is a robust least squares fit to the range $\tau_{250} = $ 0.0025--0.0065, with the slope, $k$, given at the bottom of the frame. The zero point of the $I_{\nu}$(3.6$\mu$m) axis is arbitrary.
}
\label{fig:Sp_plot}
\end{figure}

The Spitzer 3.6\,$\mu$m map and the area used in the correlations are shown in Fig.~\ref{fig:Sp_map}. The resulting correlations between MIR surface brightness and optical depth $\tau_{250}$ are shown in Fig.~\ref{fig:Sp_plot}. In the figure, we also show the median surface brightness that is calculated for $\tau_{250}$ bins with a width of 0.002. We fit a robust least squares line to the individual surface brightness values in pixels where $\tau_{250}$ is between 0.0025--0.0065. The fit is performed iteratively, discarding points falling further than 2.5-$\sigma$ from the fitted line. The value of the fitted slope is 2.40 MJy/sr.

\subsection{Spectral energy distributions} \label{sect:SED}

\begin{figure}
\centering
\includegraphics[width=9cm]{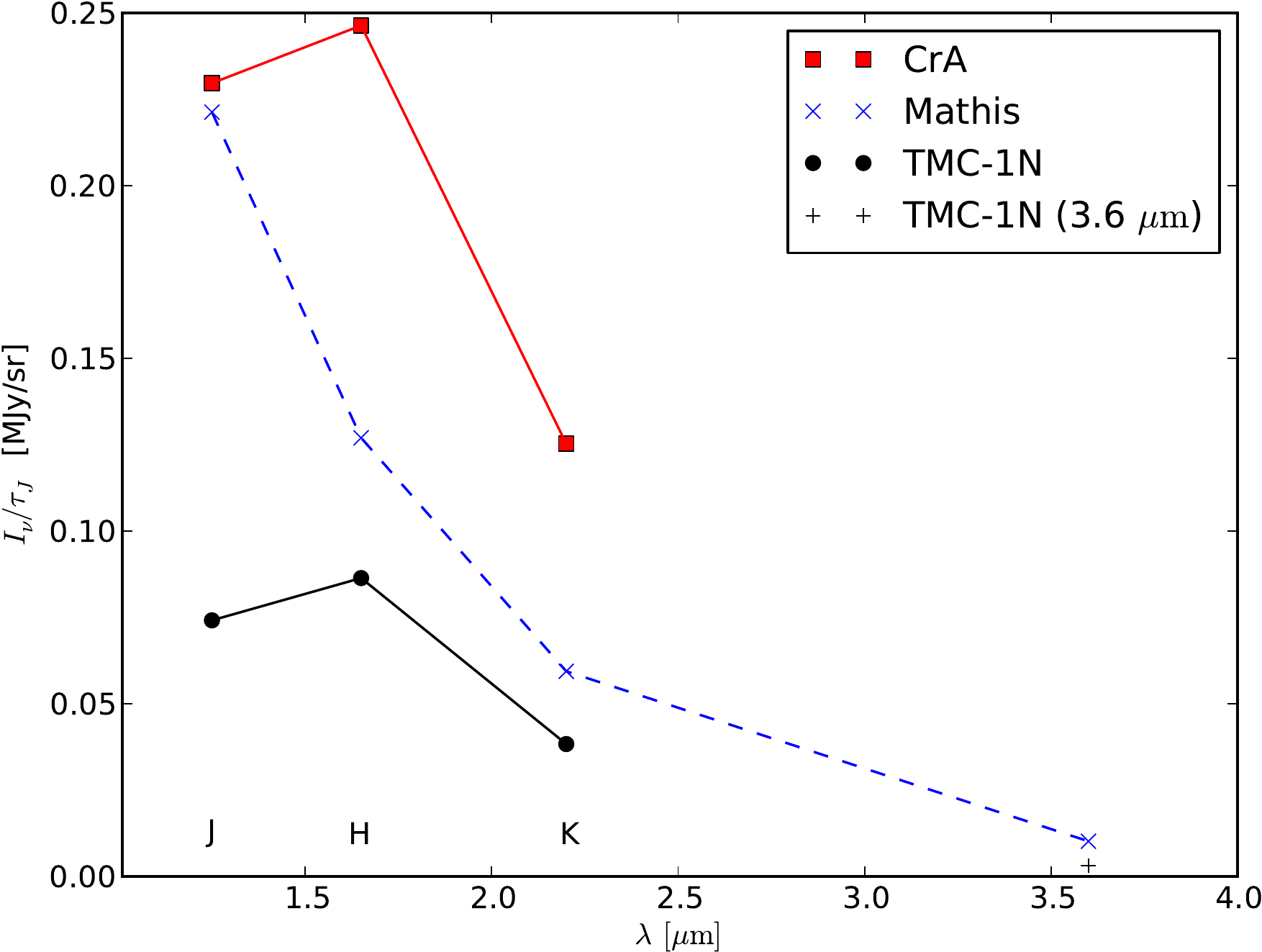}
\caption{
Spectral energy distribution shown as $I_{\nu}/\tau_J$. J, H, and K band data are shown with black circles (TMC-1N) and red squares (Corona Australis), and Spitzer 3.6 $\mu m$ data with a black plus sign. Mathis ISRF model~\citep{Mathis1983} values are marked with blue crosses and a dashed line.
}
\label{fig:SED}
\end{figure}

Spectral energy distributions are shown in Fig.~\ref{fig:SED}.
We calculate the values for $I_{\nu}/\tau_J$ in TMC-1N with Eq.~\ref{eq:I_N_lin}, using $\tau_J$ instead of column density $N$. In areas of low optical depth, the product $ab$ gives the ratio $I_{\nu}/\tau_J$. The obtained values for $J$, $H$ and $K$ bands are 0.074, 0.086, and 0.038 MJy/sr, respectively.

We compare our results with the values obtained for Corona Australis in~\citet{Juvela2008} (filled squares in their Fig. 13). We have scaled the Corona Australis values to $I_{\nu}/\tau_J$ units (with $R_V = 3.1$). We have also marked Mathis ISRF (Interstellar Radiation Field) model~\citep{Mathis1983} values (with $R_V = 4.0$) on the figure for comparison. The values are obtained by multiplying the Mathis intensities with the scattering cross-sections of the~\citet{Draine2003} dust model.

The values of $J$, $H$ and $K$ bands in TMC-1N are approximately one third of the values in Corona Australis.
Mathis model values for $H$ and $K$ are $\sim$1.5 times as high as in TMC-1N. For $J$ band, the Mathis model value is already $\sim$3 times as high as the TMC-1N values. The value we obtain for Spitzer 3.6\,$\mu$m is also only approximately one third of the value given by the Mathis model.
In both TMC-1N and Corona Australis, $I_{\nu}/\tau_J$ is lower in the $J$ band than in the $H$ band. However, in the Mathis model, the value for the $J$ band is notably higher than for the $H$ band.

\section{Radiative transfer modelling} \label{sect:models}

We carry out radiative transfer calculations of the ISRF light that is
scattered from the cloud in the wavelength range 1.2--3.5\,$\mu$m. We
construct a realistic three-dimensional model of the density
distribution of the densest part of the filament. Based on the density structure, we calculate predictions for the surface brightness using the Diffuse Infrared Background Experiment (DIRBE) observations on the \emph{Cosmic Background Explorer} (COBE) as a template of the sky brightness. The radiative transfer calculations were carried out with a Monte Carlo radiative transfer program~\citep{Juvela2003}.

The filament is illuminated by an anisotropic radiation field that, in
conjunction with the scattering phase function, affects the strength
and spatial distribution of the scattered intensity. We define the
intensity of the incoming radiation using the DIRBE all-sky
maps\footnote{See DIRBE explanatory supplement,
\tt http://lambda.gsfc.nasa.gov/product/cobe/dirbe\_exsup.cfm} (Zodi-Subtracted Mission Average (ZSMA) Maps). The DIRBE
bands 1, 2, and 3 can be used directly to specify the intensity at the
corresponding wavelengths 1.2\,$\mu$m, 2.2\,$\mu$m, and 3.5\,$\mu$m.
The $H$ band intensity is obtained by multiplying the average of the $J$
and $K$ band intensities with the intensity ratio $I_{\rm H}/<I_{\rm J},
I_{\rm K}>$ taken from the \citet{Mathis1983} ISRF model. Note that,
averaged over the whole sky, the total intensities themselves are
$\sim$50\% higher than in the \citet{Mathis1983} model.

The density distribution of the model cloud is based on the column
density map that is derived from the \emph{Herschel} sub-millimetre
observations~\citep[see][]{Malinen2012}.
This is based on the colour temperature of the
emission and on the assumption that dust opacity follows the law
0.1\,cm$^2$/g\,($\nu$/1000\,GHz)$^{\beta}$, with $\beta=2.0$. 
The actual cloud model corresponds to
$17\arcmin \times 17\arcmin$ area that we cover with 68$\times$68 pixels, 15$\arcsec$ in size.
Our three-dimensional cloud model is correspondingly a cube that consists
of $68^3$ cells and is viewed along one of its major axes.

The \emph{Herschel} column density map constrains the mass distribution only
in the plane of the sky. To construct a three-dimensional filament, we
start with a cylindrical structure where the radial density profile
follows the Plummer function
\begin{equation}
\rho_{p}(r) = \frac{\rho_{c}}{[1+(r/R_{\rm flat})^2]^{p/2}}.
\end{equation}
with parameters $\rho_{\rm C}=4\times 10^4$\,cm$^{-3}$, $R_{\rm flat}$=0.03, and $p$=3.0. These are close to the values previously
determined from the fitting of the \emph{Herschel} observations
\citep{Malinen2012}. This initial cylindrical structure therefore has
properties close to those of the average filament. We modify this
initial model by requiring that, for each line-of-sight, the column
density of the model exactly matches that of the \emph{Herschel} column
density map. We calculate for each pixel the ratio of the \emph{Herschel}
column density and the initial column density in the model. The
densities along the same line-of-sight are then multiplied by this
value. The procedure modifies the filament so that it is no longer
cylinder symmetric. However, the deformations remain relatively small
so that the ratio of the major and minor axis of the 2D filament cross
sections always remains below two.

The absolute value of the background affects the contrast between the cloud and the background.
The surface brightness relative to the brightness of the background sky is calculated as
\begin{equation}
I = I_{sca} + I_{bg}e^{-\tau} - I_{bg},
\label{eq:Ibg}
\end{equation}
where $I$ is the total observed signal, $I_{sca}$ the scattered signal coming from the cloud, $I_{bg}e^{-\tau}$ the background signal coming directly (without scattering) through the cloud with optical depth $\tau$, and $I_{bg}$ the background signal. The radiative transfer program takes into account the incoming ISRF and the absorption and scattering of those photons inside the cloud. $I_{sca}$ is the amount of light coming out of the cloud after this process.

\begin{figure}
\centering
\includegraphics[width=9cm]{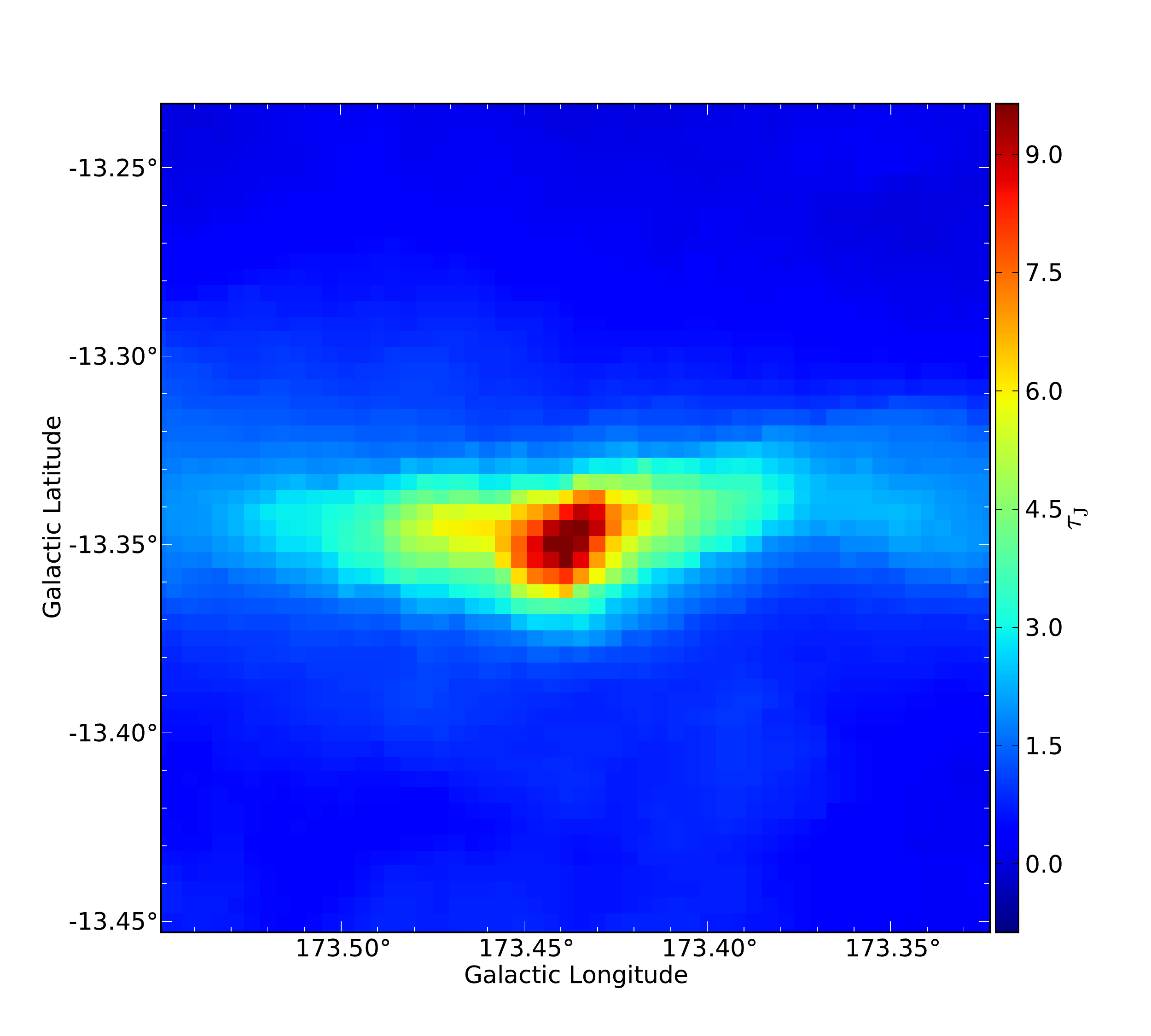}
\caption{
Optical depth $\tau_J$ of the model cloud.
}
\label{fig:simu_tauJ}
\end{figure}

The term $I_{bg}$ was estimated using WISE~\citep{Wright2010} and DIRBE data. The WISE map of a one degree diameter area around the column density peak was first adjusted to the absolute level indicated by the DIRBE data. We read the $I_{bg}$ value from the corrected WISE map, using the area that corresponds to the area where the model column density is approximately zero (corresponding to the upper part of Fig.~\ref{fig:simu_tauJ}). To exclude the effect of point sources in this area, we use the 40\% percentile of pixels in this reference area.
The values obtained for the background $I_{bg}$ are 0.122 MJy/sr ($J$), 0.0878 MJy/sr ($K$), and 0.0765 MJy/sr (3.4\,$\mu$m). We estimate a value 0.10 MJy/sr for the $H$ band. 
The cosmic infrared background (CIB) between the $J$ and 3.5$\mu$m bands is notably lower than these values, $\sim$0.01--0.02 MJy/sr~\citep[see, e.g.,][]{Cambresy2001,Wright2000}.

The dispersion of DIRBE values over the area used for comparison with WFCAM and WISE is $\sim$10\% of the estimated $I_{bg}$ value. Because this includes not only noise but also real surface brightness variations, the statistical error of the mean DIRBE value is significantly lower. 
However, the dispersion is also a measure of the
possible difference between the background value $I_{bg}$ derived for the reference area and the actual background at the position of the filament.
The final $I_{bg}$ is calculated as 40\% percentile of the reference area within WFCAM and WISE maps. The standard deviations of all pixels falling below the 40\% value (assumed not to be significantly contaminated by point sources) are 0.011, 0.015, and 0.005\,MJy\,sr$^{-1}$ for $J$, $K$, and 3.4\,$\mu$m bands, respectively. The zero point uncertainty of the zodiacal light model subtracted in DIRBE ZSMA maps is $\sim$0.006\,MJy\,sr$^{-1}$ in the $J$ band and less at longer wavelengths \citep{Kelsall1998}, but differences between zodiacal light models can be even larger \citep{Wright2001}. Furthermore, Taurus is located at low ecliptic latitude and two thirds of the total signal observed by DIRBE consists of zodiacal light. A relative error of 10\% in the zodiacal light model would therefore amount to $\sim$0.04, 0.03, and 0.02\,MJy\,sr$^{-1}$ for $J$, $K$, and 3.5\,$\mu$m, respectively, and could be the dominant source of error.

Diffuse emission that originates between the filament and observer does not contribute to Eq.~\ref{eq:Ibg}, which describes the surface brightness contrast between the source and the background. In that equation, $I_{bg}$ corresponds to that part of the diffuse component that truly resides behind the filament. We can assume that most of the diffuse material is associated with the Taurus cloud but can reside either in front of or behind the filament. We therefore make the assumption that $I_{bg}$ corresponds to half of the values derived above.

If the diffuse component is closely associated with the filament, consistency requires that its effect is also taken into account when calculating the component $I_{sca}$. This term corresponds to photons scattered from the filament, excluding photons that may be scattered in some envelope around it. In this case, the filament (i.e., the densest part that is being modelled) is not illuminated by the full ISRF but by an ISRF that is attenuated by $e^{-0.5\tau_d}$, where $\tau_d$ is the optical depth of the envelope. Similarly, when a photon scatters within the model, it must travel through a similar layer a second time (half of the full $\tau_d$) before reaching the observer. Thus, the observed signal is reduced by a factor $e^{-\tau_d}$.

We estimate that for the diffuse cloud $\tau^J_d = 0.5$, and calculate values 0.34, 0.2, and 0.0998 for $\tau^H_d$, $\tau^K_d$, and $\tau^{3.5\mu \rm m}_d$, respectively, using~\citet{Cardelli1989} extinction curves.
We compare three cases: (1) no background $I_{bg}$, (2) with background $I_{bg}$, and (3) with background $I_{bg}$ and attenuation of the scattered light in the envelope. We show the corrections in function
\begin{equation}
I = I_{sca}e^{-\tau_d} + CI_{bg}(e^{-\tau} - 1),
\label{eq:Ibg_corr}
\end{equation}
where $C$ is either 0 (case 1) or 0.5 (cases 2 and 3), indicating the amount of background, and use value 0 for $\tau_d$ for cases 1 and 2, and the $\tau_d$ values derived above for case 3.

In the calculations, we use the NIR dust properties of normal Milky Way dust~\citep{Draine2003} and the tabulated scattering phase functions that are publicly available\footnote{\tt http://www.astro.princeton.edu/$\sim$draine/dust/scat.html}.
We do not seek a perfect match to the observations but explicitly
assume the column density distribution and dust properties as
described above. The comparison with the observations is thus a test for the consistency of
these assumptions.

\begin{figure*}
\centering
\includegraphics[width=16cm]{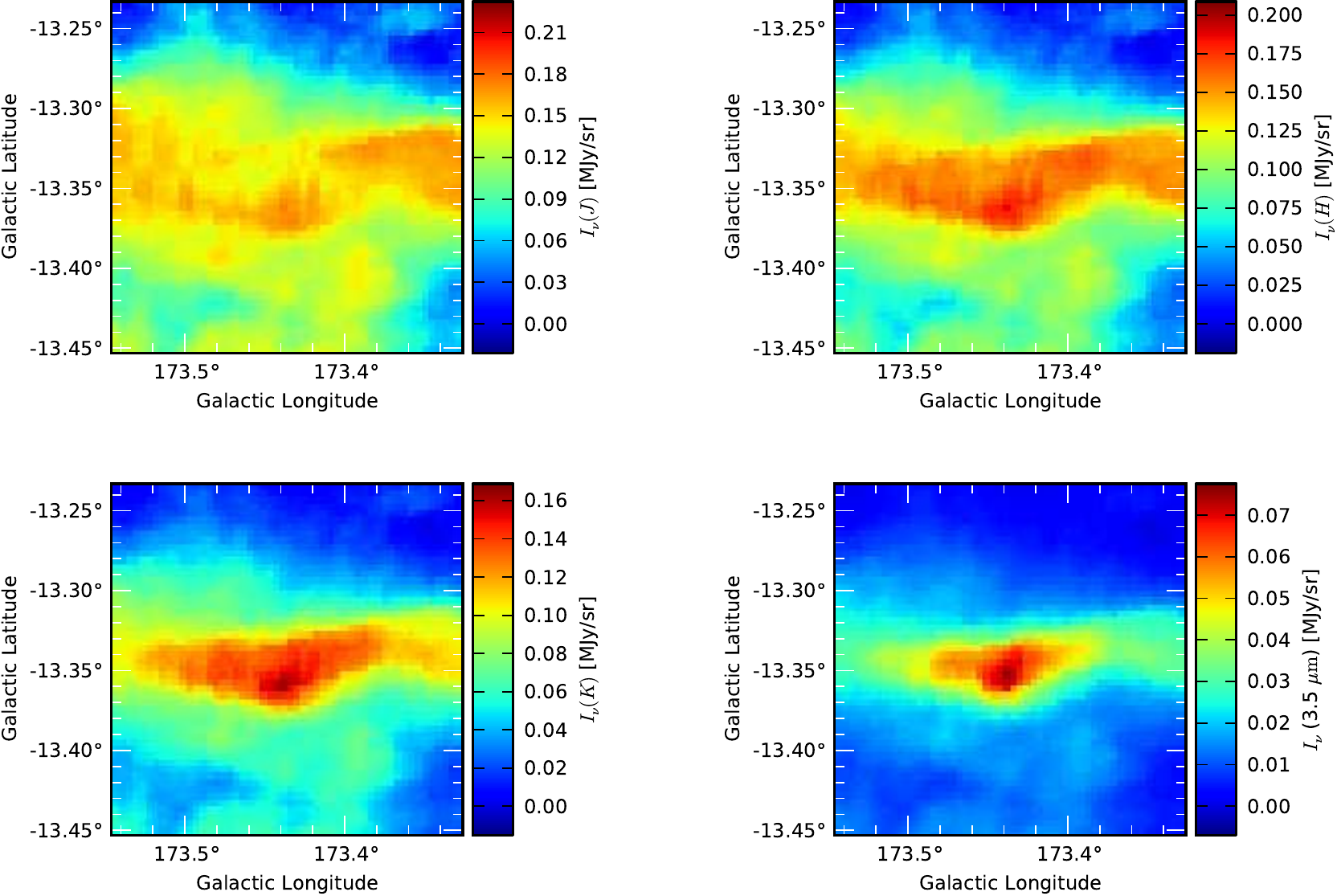}
\caption{
Simulated maps of scattered surface brightness in $J$, $H$, $K$, and 3.5\,$\mu$m bands. 
}
\label{fig:simu_map2}
\end{figure*}

\begin{figure*}
\centering
\includegraphics[width=6cm]{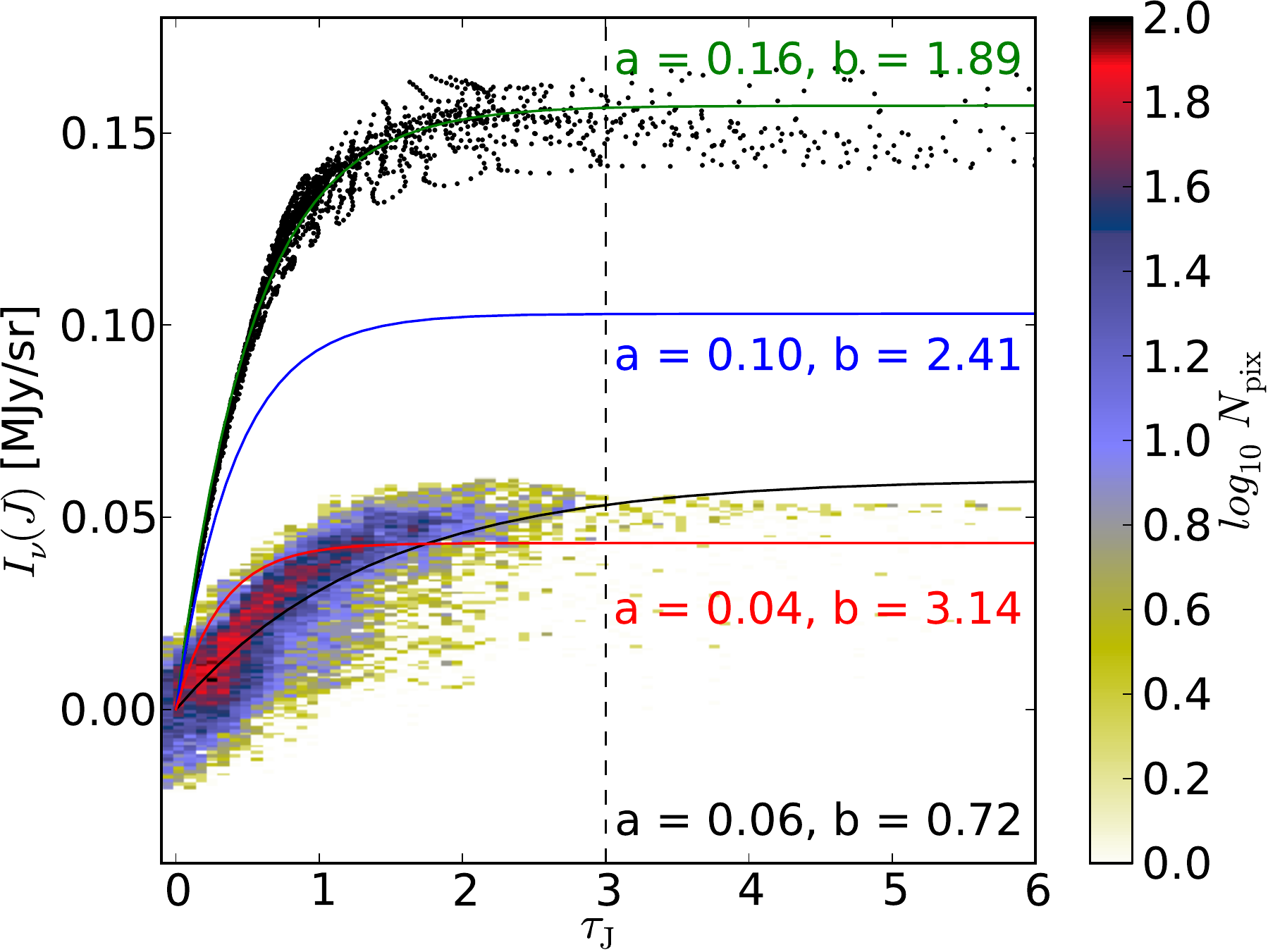}
\includegraphics[width=6cm]{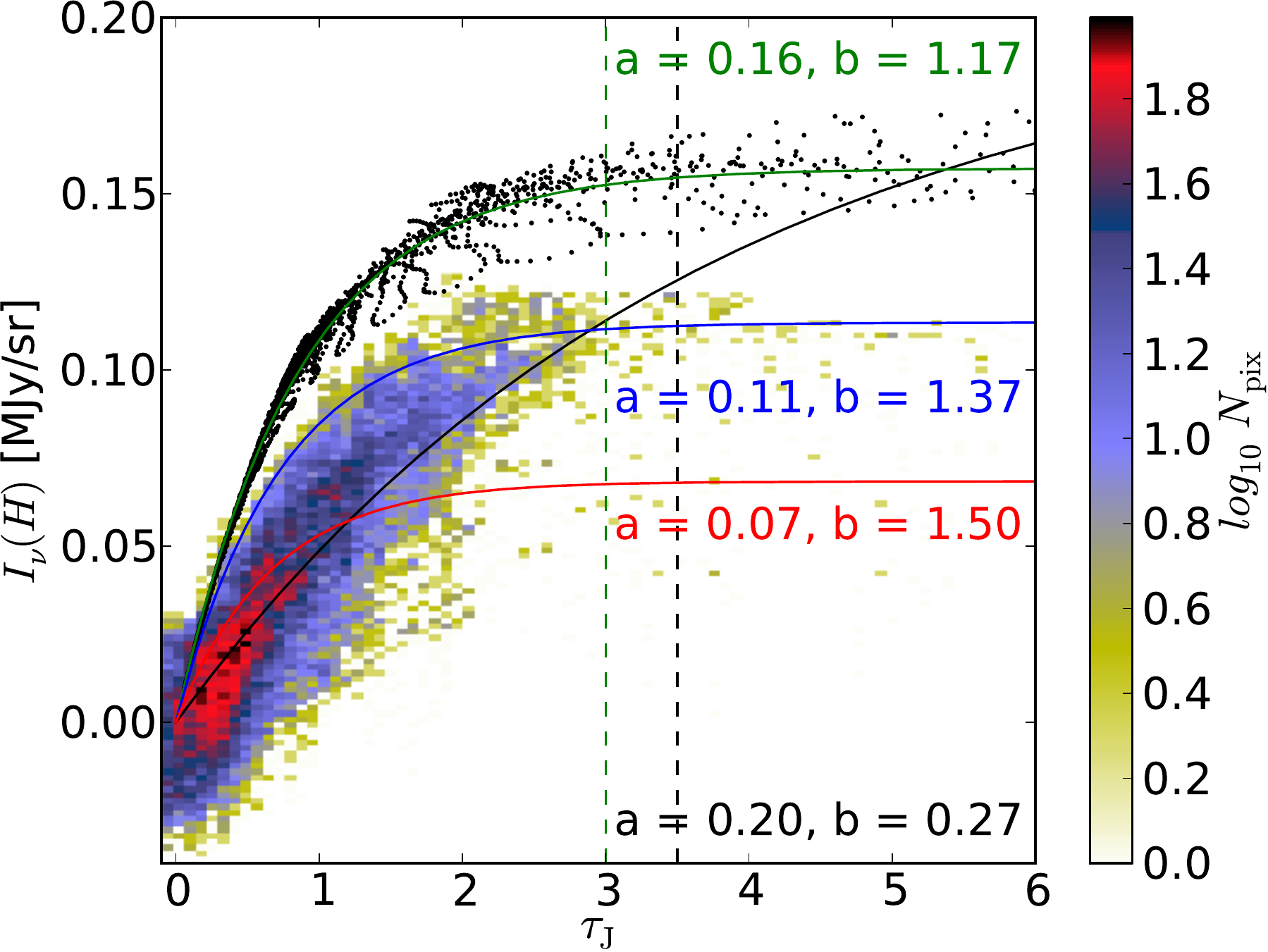}
\includegraphics[width=6cm]{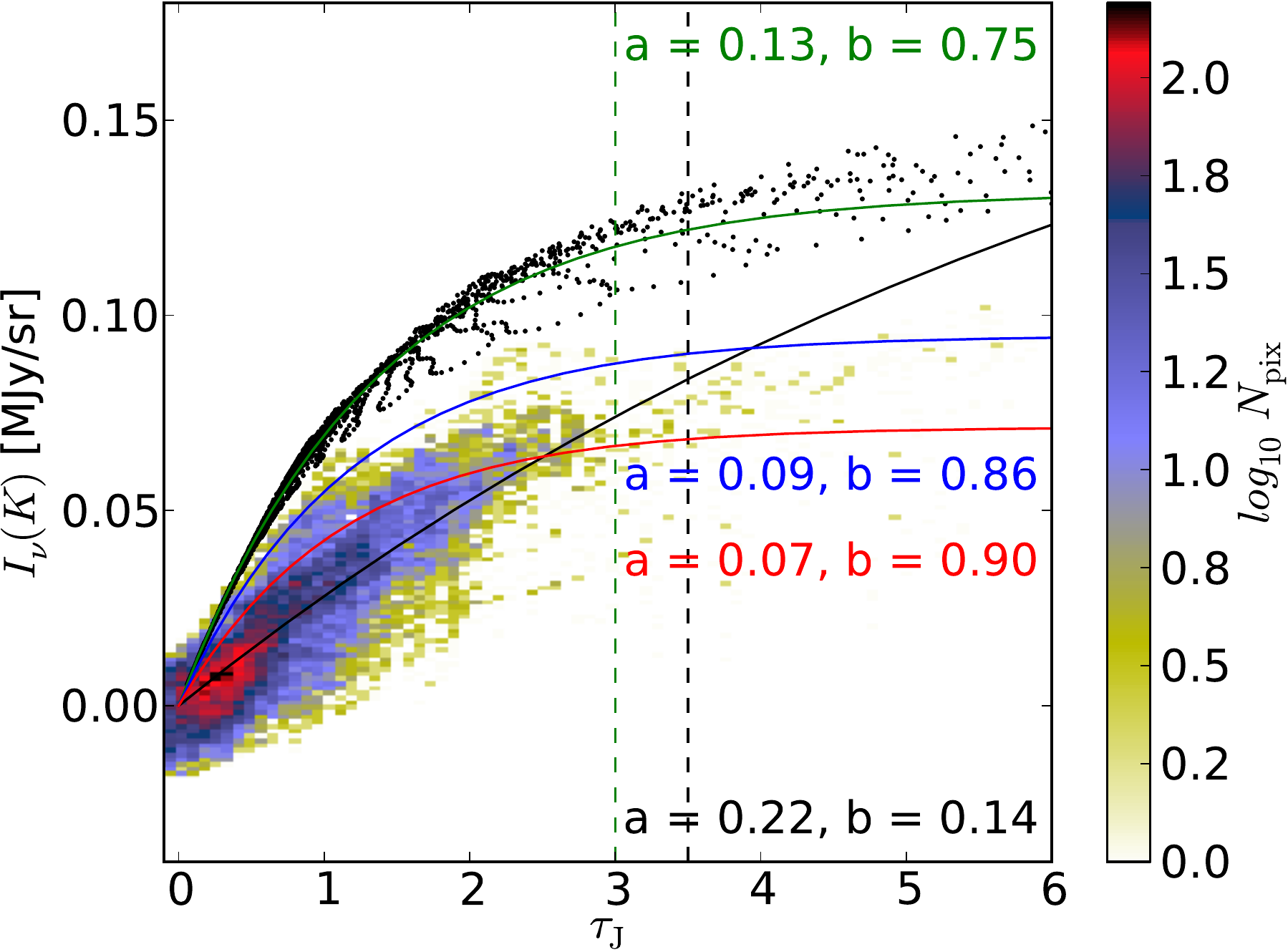}
\caption{
Comparison of observed and simulated NIR surface brightness $I_{\nu}$ in $J$, $H$, and $K$ bands as the function of optical depth in $J$ band, $\tau_{J}$. WFCAM data for TMC-1N are shown with a 2D histogram, the colour scale corresponding to the density of points, and a black line shows the fitted Eq.~\ref{eq:I_N}. For comparison, simulated data are shown in the same figures using the three test cases described in the text. We plot the simulated data points only for case 1 (black dots).
The fitted function is marked with a green line (case 1: without background), blue line (case 2: with background), and red line (case 3: with background and attenuation of the scattered light in the envelope).
In $J$ band, black dashed vertical line shows the upper limit of fitting of both WFCAM and simulation data. In $H$ and $K$ bands, the black dashed vertical line shows the upper limit of fitting of WFCAM data, and green dashed vertical line the upper limit of fitting of simulated data (all three cases). To get a better fit, the limit for the observations is slightly higher.
The obtained parameter values are marked in the figures with the same colours as the fitted lines.
}
\label{fig:simu_plot}
\end{figure*}

The $\tau_J$ map of the model is shown in Fig.~\ref{fig:simu_tauJ} and simulated scattered surface brightness maps in Fig.~\ref{fig:simu_map2}. Correlations of $I_{\nu}$ and $\tau_{J}$ from simulations are shown in Fig.~\ref{fig:simu_plot} for intensity in bands $J$, $H$, and $K$, using the three cases for the background correction described above. For comparison, we plot in the same figures the data from our TMC-1N observations. We have subtracted the background from both the simulated and observed data using as reference area the area in which the model column density is lowest (corresponding to the upper part of Fig.~\ref{fig:simu_tauJ}). In the WFCAM data, the median $\tau_J$ in this area is $\sim$0.5.
Note the difference from Fig.~\ref{fig:scpl_tauJ_I}, in which we used as reference area the low optical depth area marked in Fig.~\ref{fig:tau_fullmaps}. We fit Eq.~\ref{eq:I_N} to both of the data, using optical depth $\tau_J^{Nicer}$ instead of column density $N$ in the exponential part. As a result, we obtain the $a$ and $b$ parameters for each band.

In the $H$ band, the simulations give $\sim$1.5 times as high intensity values as the observations, without background (case 1). The observed values are mostly settled between the cases 2 and 3 in the simulations.
In the $K$ band, the simulations give nearly two times as high values to the observations, when no background is included (case 1). In case 3, with the background and attenuation in the envelope, the simulations give similar values as the observations.
However, in the $J$ band, the simulated values are approximately three times as high as in our observations, when no background is included (case 1). Even with background (case 2), the simulations still give approximately two times as high values. Only in case 3, with the attenuation in the envelope, the simulations give similar values to the observations. Even when the maximum intensities for observations and simulations are approximately the same, the form of the fitted Eq.~\ref{eq:I_N} can be different, leading to notably different values for the parameters.

\begin{figure}
\centering
\includegraphics[width=9cm]{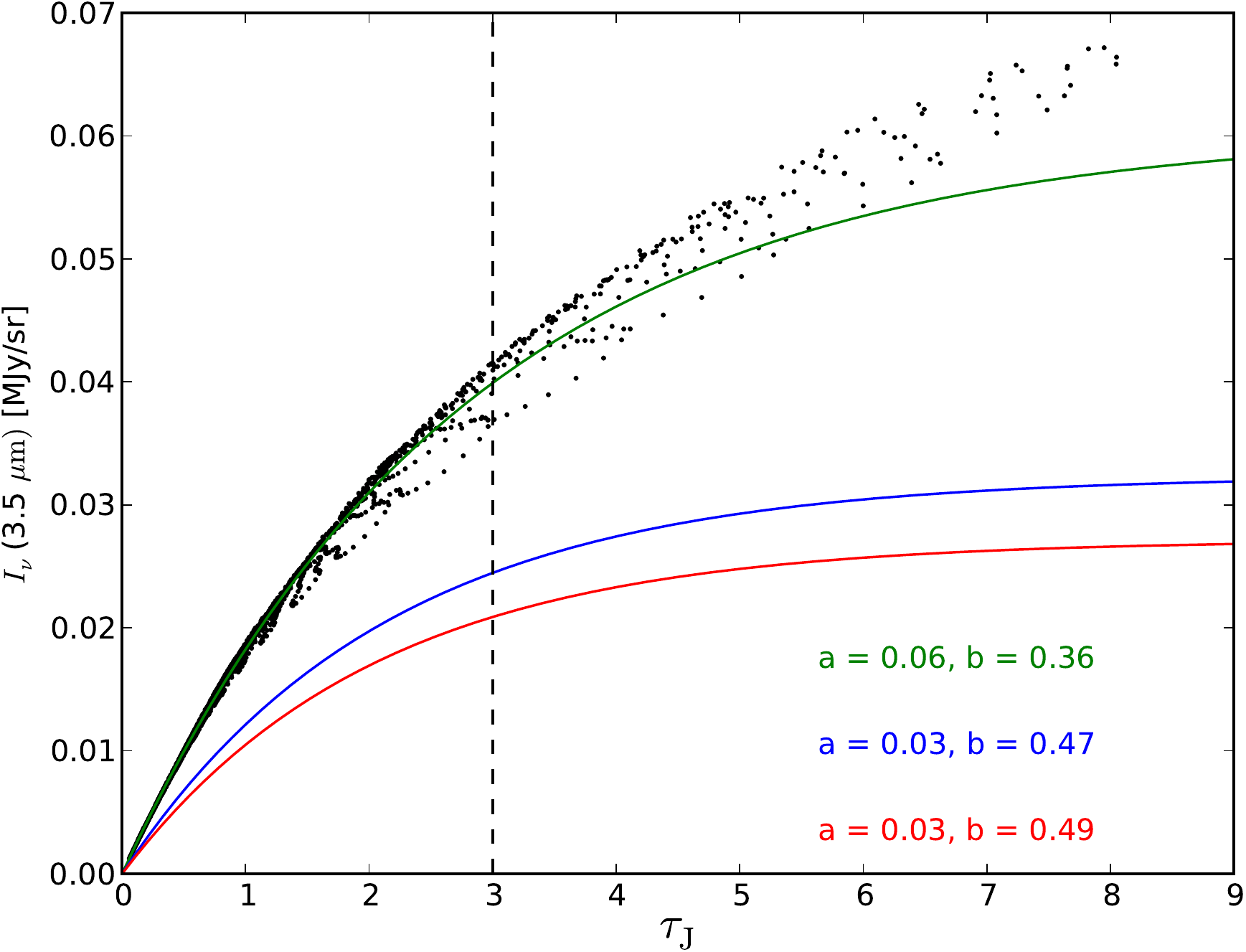}
\caption{
Simulated MIR surface brightness $I_{\nu}$ in 3.5\,$\mu$m band as a function of $\tau_{J}$ in the three test cases described in the text. The data are fitted with Eq.~\ref{eq:I_N}, and marked with the same notation as in Fig.~\ref{fig:simu_plot}. The black dashed line shows the upper limit for the data used in the fit. The obtained parameter values are marked in the figure.
}
\label{fig:simu_plot_3.5}
\end{figure}

Correlation between $I_{\nu}$ and $\tau_{J}$ for the 3.5\,$\mu$m band is shown in Fig.~\ref{fig:simu_plot_3.5}, similarly using the three cases. At this wavelength, case 1 still gives two times as high values as case 2, but the difference between cases 2 and 3 is only $\sim$10\%.

\section{Discussion} \label{sect:discussion}

We have studied a filament in the Taurus molecular cloud using NIR images of scattered light. The observations carried out with WFCAM instrument cover an area of $1^{\circ}\times 1^{\circ}$ corresponding to $\sim$(2.44 pc)$^2$ at the distance of 140 pc, making this, to our knowledge, 
the largest NIR map where the surface brightness is analysed in detail.
We have analysed the faint surface brightness, determined its intensity (attributed to light scattering), and used it to derive an optical depth map based on the method of \citet{Padoan2006}. We have compared the data derived from NIR scattering, NIR extinction and \emph{Herschel} dust emission.

The signal-to-noise value, S/N, of our WFCAM data is lower than expected. There were significant artefacts, presumably of instrumental origin, namely the curtain effect and large scale gradients. It is possible that the excess noise is caused by these effects and the processes used to remove them. Due to this, the data could not be used at as high a resolution as expected. In principle, scattered light can be measured at arcsecond resolution. However, as shown here, it can be difficult to obtain reliable maps at the full resolution.

Correlations between $I_{\nu}$ and $\tau_{J}^{Nicer}$ in Fig.~\ref{fig:scpl_tauJ_I} suggest that the radiation field in TMC-1N is notably lower in all three NIR bands when compared to similar observations of Corona Australis~\citep{Juvela2008}. This reaffirms the results obtained with dust emission models of \citet{Juvela2012b}, suggesting that the radiation field in Corona Australis is at least three times that of the normal ISRF.

NIR intensity as a function of column density followed roughly the functional form assumed in Eq.~\ref{eq:I_N}.
We have estimated the possible errors in obtaining the parameters $a$ and $b$ for each band from fitting Eq.~\ref{eq:I_N} to the correlation between $I_{\nu}$ and $\tau_{J}^{Nicer}$. Errors caused by sampling are small, but changes to the fitting limit can cause up to $\sim$60\% deviations in a single parameter. However, the change in the product $a \times b$, that is the slope of the linear part of the Eq.~\ref{eq:I_N}, is only up to $\sim$10\% for each band.
As discussed in Sect.~\ref{sect:method}, Eq.~\ref{eq:I_N} and the parameters $a$ and $b$ can be thought of as an empirical model that represents the relation between surface brightness and column density. The model may not be optimal and no direct physical interpretation can necessarily be attached to its parameters.
Bias in the fitted parameter values could cause bias also when minimising the Eq.~\ref{eq:tauSB_1} to obtain an optical depth map.

We obtained the value 0.0013 for the slope $\tau_{250}/\tau_{J}^{Nicer}$ with $\beta = 1.8$, using values in the range $\tau_{J}=$0--2 (and the same for range 0--4), in Fig.~\ref{fig:scpl_tauNJ_tau250}. 
When we fit data in the range 2--4, the slope increased $\sim$15\% to 0.0015, possibly suggesting a small increase in grain size in the denser areas, similarly to~\citep{Martin2012,Roy2013}. With $\beta = 2.0$, and using range 0--4, the slope increased $\sim$38\% to 0.0018.

For comparison with other studies, we convert our slope of $\tau_{250}/\tau_{J}^{Nicer} = 0.0013$ to dust absorption cross-section per H nucleon, also called opacity. We use the conversion factor $N({\rm HI} + {\rm H}_2)/E(B-V) = 5.8 \times 10^{21}$ cm$^{-2}$/mag of \citet{Bohlin1978}. In principle, this relation is valid only for diffuse areas, and can be used only as an approximation for the densest parts of our filament. We derive the extinction relations, $E(B-V)/E(J-K)$ and $A_J/E(J-K)$, using the extinction curves of \citet{Cardelli1989}, either with $R_V = 3.1$ or $R_V = 4.0$. This leads to $\sigma_e(250 \mu \rm{m})$ values $1.7 \times 10^{-25} {\rm cm}^2/{\rm H}$ (with $R_V = $3.1) or $2.4 \times 10^{-25} {\rm cm}^2/{\rm H}$ (with $R_V = 4.0$). In terms of mass absorption (or emission) coefficient per gas mass, $\kappa_{\nu}$, the values are 0.07 cm$^2$/g or 0.10 cm$^2$/g, respectively. Changing the assumed $R_V$ value in $E(B-V)/E(J-K)$ from 3.1 to 4.0 causes $\sim$40\% increase in the obtained values. It is not obvious which assumptions can be made in areas having both diffuse and denser parts, but it is important to notice that the assumptions made will have notable differences in the results.

For high-latitude, diffuse ISM, the standard value for dust opacity is $\sigma_e(250\mu \rm{m})\sim1.0\times10^{-25} {\rm cm}^2/{\rm H}$ \citep[see, e.g.,][]{Boulanger1996,Planck2011b}. In denser areas, 2--4 times larger values have been derived~\citep[see, e.g.,][]{Juvela2011,Planck2011b,Martin2012,Roy2013}. Our estimates for $\sigma_e(250 \mu \rm{m})$ are $\sim1.7-2.4$ times as large as the previous results for diffuse areas, at the lower limit of the values for denser areas.

As the width (or FWHM) of our filament is $\sim$0.1 pc, which for Taurus corresponds to $\sim$150$''$,
we have enough resolution to look at the details of the filament. The $\tau_{250}/\tau_{J}^{Nicer}$ map shown in Fig.~\ref{fig:map_tau250_I} gives no indication of systematic growth of the $\tau_{250}/\tau_{J}^{Nicer}$ towards the densest filament. However, as discussed above, all the assumptions made in the process may not be valid in the densest part of the filament. As an additional confusing factor, some areas of high $\tau_{250}/\tau_{J}^{Nicer}$ ratio can be caused by small differences in the $\tau_{250}$ and $\tau_{J}^{Nicer}$ maps, due to, e.g., the lack of background stars seen behind the densest filament, instead of real changes in the dust properties.

In Sect.~\ref{sect:models}, we constructed a realistic three-dimensional model of the density distribution of the densest part of the filament. We calculated predictions for the surface brightness using radiative transfer calculations of the ISRF light that is scattered from the cloud in the wavelength range 1.2--3.5\,$\mu$m.
We used DIRBE observations as a model for the sky brightness. 
As the absolute value of the background affects the contrast between the cloud and the background, we calibrated our simulation data to the same absolute level with DIRBE. We also 
considered the role of a potential diffuse envelope that would affect the radiation impinging on the dense part of the filament, i.e., the part included in our model. We take into account that a scattered photon has to pass through the diffuse cloud twice. We compared three cases of using different background correction: (1) no background, (2) with background, and (3) with background and attenuation of the scattered light in the envelope surrounding the filament.

In the simulations, we have assumed that the filament is in the plane of the sky and is not prolate (long along the line-of-sight).
If the filament is prolate, this will increase the short wavelength surface brightness relative to the longer wavelengths, and the saturation of the $J$ band will be smaller.
If the filament is not in the plane of the sky (i.e., our line-of-sight is not perpendicular to the filament axis), the effect is similar.
The dust opacities used in the simulations are appropriate in high density environments \citep{Hildebrand1983,Beckwith1990} but could, in diffuse regions, overestimate $\kappa$ (underestimate column density) by a factor of two \citep[see][]{Boulanger1996}.

We find that the results change notably, especially for the $J$ band, depending on what fraction of the background intensity is assumed to be really behind the filament and what fraction between the filament and the observer. In the $J$ band, case 1 gives $\sim$1.5 times as high values to the relation between $I_{\nu}$ and $\tau_{J}$ as case 2, and $\sim$3 times as high values as case 3. In longer wavelengths the differences get smaller, but at 3.5\,$\mu$m, case 1 still gives $\sim$2 times as high values as case 2. At 3.5\,$\mu$m, the difference between cases 2 and 3 is only $\sim$10\%.

In the $H$ and $K$ bands, the simulations give rather similar results to our observations in TMC-1N. However, in the $J$ band, the simulations give over two times as high values as the observations, unless case 3 with the background correction and the correction for the scattered light is used. In the $H$ band, case 3 also gives values that are lower than the observed values. The three test cases can be seen as rough error limits for the modelled cloud.
Compared to the simulations, our observations do not suggest that there is any notable emission in the $K$ band in addition to the scattered light.

\section{Conclusions} \label{sect:conclusions}

We have used WFCAM NIR surface brightness observations to study scattered light in the TMC-1N filament in Taurus Molecular Cloud.
We have presented a large NIR surface brightness map ($1^{\circ} \times 1^{\circ}$ corresponding to $\sim$(2.44 pc)$^2$) of this filament.
We have converted the data into an optical depth map and compared the results with NIR extinction and \emph{Herschel} observations of sub-mm dust emission. We have also modelled the filament by carrying out 3D radiative transfer calculations of light scattering.

\begin{itemize}

\item We see the filament in scattered light in all three NIR bands, $J$, $H$, and $K$.

\item In all three NIR bands, our WFCAM observations in TMC-1N show lower intensity than previous results in Corona Australis, indicating a lower radiation field in this area. This reaffirms the previous findings, that the radiation field in Corona Australis is at least three times that of the normal ISRF.

\item We derive a value 0.0013 for the ratio $\tau_{250}/\tau_{J}^{Nicer}$ . This leads to values $\sigma_e(250 \mu \rm{m})\sim1.7-2.4\times10^{-25} {\rm cm}^2/{\rm H}$, depending on the assumptions of the extinction curve ($R_V = $ 3.1 or 4.0) which can change the results by over 40\%. These $\sigma_e(250 \mu \rm{m})$ values are twice the values reported for diffuse medium, at the lower limit of the values for denser areas.

\item Changing $\beta$ from 1.8 to 2.0 increases the ratio $\tau_{250}/\tau_{J}^{Nicer}$ by $\sim$30\%.

\item We see no indication of systematic growth of the $\tau_{250}/\tau_{J}^{Nicer}$ ratio towards the densest filament. However, all the assumptions made in the process may not be valid in the densest part of the filament. Also, some areas of high $\tau_{250}/\tau_{J}^{Nicer}$ ratio can be caused by imperfections in the $\tau_{250}$ and $\tau_{J}^{Nicer}$ maps, due to, e.g., the lack of background stars seen behind the densest filament, instead of real changes in the dust properties.

\item 3D radiative transfer simulations predict surface brightness that is in intensity close to the observed values, especially in the $H$ and $K$ bands. In the $J$ band, the model predictions can be over two times larger than observations, if no background correction is made. However, using background correction can change the results notably.

\item We see no clear evidence for emission in the $K$ band, in addition to the scattered light, based on the observations and simulations.

\item NIR surface brightness can be a valuable tool in making high resolution maps, also at large scales.

\item NIR surface brightness observations can be complicated, however, as the data can show comparatively low-level artefacts, that are still comparable to the faint surface brightness. This suggests caution when planning and interpreting the observations.

\item It is possible to remove most of the effects of instrumental gradients, provided that they only affect large scales.

\end{itemize}

\acknowledgements
We thank the referee for useful comments. We thank CASA for carrying out the standard data reduction of the WFCAM data, and Mike Irwin for useful comments.
JM and MJ acknowledge the support of the Academy of Finland Grants No. 250741 and 127015.
MGR gratefully acknowledges support from the National Radio Astronomy Observatory (NRAO), the Joint ALMA Observatory and the Joint Astronomy Centre, Hawaii (UKIRT). The National Radio Astronomy Observatory is a facility of the National Science Foundation operated under cooperative agreement by Associated Universities, Inc.

The United Kingdom Infrared Telescope is operated by the Joint Astronomy Centre on behalf of the Science and Technology Facilities Council of the U.K.
This paper makes use of WFCAM observations processed by the Cambridge
Astronomy Survey Unit (CASU) at the Institute of Astronomy, University
of Cambridge.

This publication makes use of data products from the Two Micron All Sky Survey, which is a joint project of the University of Massachusetts and the Infrared Processing and Analysis Center/California Institute of Technology, funded by the National Aeronautics and Space Administration and the National Science Foundation.

This research made use of Montage, funded by the National Aeronautics and Space Administration's Earth Science Technology Office, Computation Technologies Project, under Cooperative Agreement Number NCC5-626 between NASA and the California Institute of Technology. Montage is maintained by the NASA/IPAC Infrared Science Archive.

This work is based in part on observations made with the Spitzer Space Telescope, which is operated by the Jet Propulsion Laboratory, California Institute of Technology under a contract with NASA.

This publication makes use of data products from the Wide-field Infrared Survey Explorer, which is a joint project of the University of California, Los Angeles, and the Jet Propulsion Laboratory/California Institute of Technology, funded by the National Aeronautics and Space Administration.

\bibliographystyle{aa}
\bibliography{biblio_j2}

\end{document}